\pdfmapline{-dummy LMRoman7-Regular}

\documentclass[a4paper, 12pt]{article}
\usepackage[utf8]{inputenc}
\usepackage[default]{lato}
\usepackage{graphicx}
\usepackage{wrapfig}
\usepackage{xcolor}
\usepackage{sectsty}
\usepackage{hyperref}
\usepackage{url}
\usepackage{breakurl}
\usepackage{amsmath}
\usepackage{multicol}
\usepackage{changepage}
\usepackage[skip=5pt]{caption}
\usepackage[style=nature,minbibnames=3, natbib=true, isbn=false, doi=true]{biblatex}
\AtEveryBibitem{\clearfield{note}}
\AtEveryBibitem{\clearfield{pages}}

\usepackage[left=2.5cm, right=2.5cm, top=2.5cm, bottom=2.5cm]{geometry}

\definecolor{bleu_LAB}{RGB}{42,57,144}
\definecolor{rouge_LAB}{RGB}{156,37,77}
\definecolor{vert_LAB}{RGB}{56,118,29}
\definecolor{gris_fonce_LAB}{RGB}{67,67,67}
\definecolor{gris_clair_LAB}{RGB}{153,153,153}

\sectionfont{\color{bleu_LAB}}
\subsectionfont{\color{rouge_LAB}}
\subsubsectionfont{\color{vert_LAB}}

\title{\vspace{+3.5cm} \color{gris_fonce_LAB} \textbf{Atmospheric Chemistry and Surface Interactions on Mars (ACSIOM) workshop} \vspace{-0.25cm}}
\author{\small {\color{gris_fonce_LAB} Benjamin Benne$^{1,2}$}, {\color{gris_fonce_LAB} Megan A. J. Brown$^{3}$}, {\color{gris_fonce_LAB} Bethan S. Gregory$^{4}$}, {\color{gris_fonce_LAB} Paul M. Streeter$^{5}$} {\color{gris_fonce_LAB}and the ACSIOM Team}}

\date{}

\begin{document}

\maketitle
\footnotesize{\noindent $^1$ School of GeoSciences, The University of Edinburgh, Edinburgh, UK \\
$^2$ Centre for Exoplanet Science, University of Edinburgh, Edinburgh, UK \\
$^3$ Centre for Atmospheric Science, Yusuf Hamied Department of Chemistry, University of\\ Cambridge, Cambridge, UK \\
$^4$ Department of Physics, University of Oxford, UK \\
$^5$ School of Physical Sciences, The Open University, Milton Keynes, UK \\}

\begin{figure}[!t]
    \begin{flushleft}
    \includegraphics[scale=1.0]{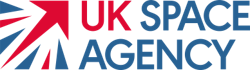}
    \end{flushleft}
\end{figure}

\color{gris_fonce_LAB}

\vspace{-1cm}
\begin{center}
\includegraphics[width=0.8\textwidth]{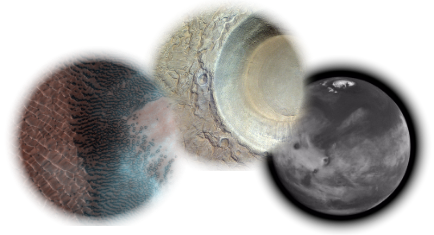}
\end{center}

\begin{figure}[!b]
    \includegraphics[width=0.3\linewidth]{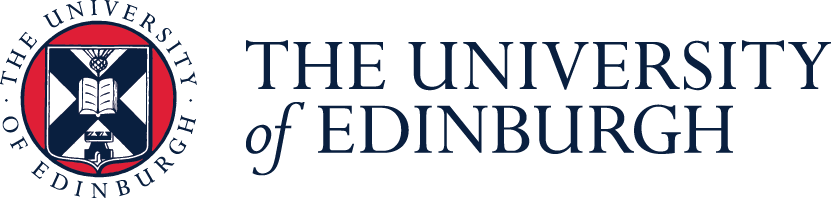}
    \hspace{0.2cm}
    \includegraphics[width=0.25\linewidth]{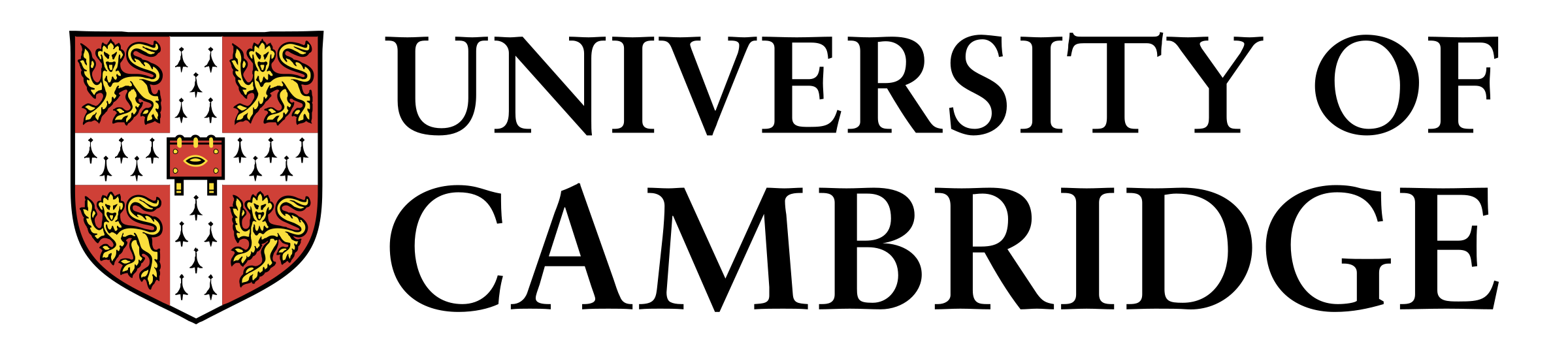}
    \hspace{0.2cm}
    \includegraphics[width=0.2\linewidth]{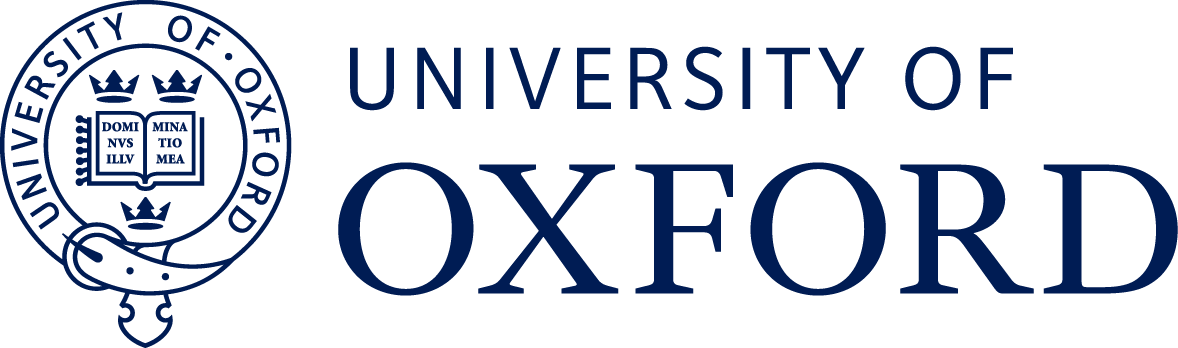}
    \hspace{0.2cm}
    \includegraphics[width=0.18\linewidth]{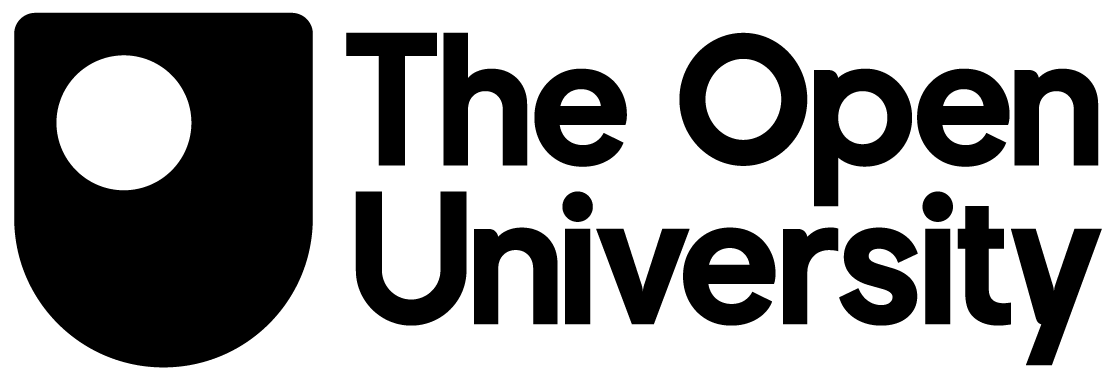}
\end{figure}

\thispagestyle{empty}
\newpage

\thispagestyle{empty}
\newgeometry{left=2cm, right=2cm, top=2.5cm, bottom=2.5cm}

\section*{Authors}

\begin{table}[h]
\begin{tabular}{llll}
{\color{gris_fonce_LAB} Benjamin Benne$^{1,2}$}  & {\color{gris_fonce_LAB} Megan A. J. Brown$^{3}$} & {\color{gris_fonce_LAB} Bethan S. Gregory$^{4}$} & {\color{gris_fonce_LAB} Paul M. Streeter$^{5}$} \\
{\color{gris_fonce_LAB} Alexander T. Archibald$^{3,6}$} & {\color{gris_fonce_LAB} Steven Banham$^{7}$}     & {\color{gris_fonce_LAB} Raluca Blasciuc$^{8}$}   & {\color{gris_fonce_LAB} Joanna Egan$^{5}$}       \\
{\color{gris_fonce_LAB} Sanjeev Gupta$^{9}$}     & {\color{gris_fonce_LAB} James A. Holmes$^{5}$}   & {\color{gris_fonce_LAB} Lorrie S. D. Jacob$^3$}  & {\color{gris_fonce_LAB} Franck Lefèvre$^{10}$}  \\
{\color{gris_fonce_LAB} Stephen R. Lewis$^{5}$}  & {\color{gris_fonce_LAB} Will Luscombe$^{11}$}    & {\color{gris_fonce_LAB} Ehouarn Millour$^{12}$}  & {\color{gris_fonce_LAB} Claire Newman$^{5,13}$} \\
{\color{gris_fonce_LAB} Kevin Olsen$^{4}$}       & {\color{gris_fonce_LAB} Paul I. Palmer$^{1,2}$}  & {\color{gris_fonce_LAB} Manish R. Patel$^{5}$}   & {\color{gris_fonce_LAB} Kylash Rajendran$^5$}   \\
{\color{gris_fonce_LAB} Paul B. Rimmer$^{14}$}   & {\color{gris_fonce_LAB} Vinayak Shastri$^{5}$}   & {\color{gris_fonce_LAB} Jon Telling$^{8}$}       & {\color{gris_fonce_LAB} Sam Webber$^{3}$}
\end{tabular}
\end{table}

~
\newline

\noindent $^1$ School of GeoSciences, The University of Edinburgh, Edinburgh, UK \\
$^2$ Centre for Exoplanet Science, University of Edinburgh, Edinburgh, UK \\
$^3$ Centre for Atmospheric Science, Yusuf Hamied Department of Chemistry, University of\\ Cambridge, Cambridge, UK \\
$^4$ Department of Physics, University of Oxford, UK \\
$^5$ School of Physical Sciences, The Open University, Milton Keynes, UK \\
$^6$ National Centre for Atmospheric Science, UK \\
$^7$ Imperial College, London, UK \\
$^8$ Newcastle University, Newcastle, UK \\
$^9$ Department of Earth Science and Engineering, Imperial College, London, UK \\
$^{10}$ LATMOS, Sorbonne Université, UVSQ Université Paris-Saclay, CNRS, Paris, France\\
$^{11}$ Department of Physics and Astronomy, University of Exeter, Exeter, UK \\
$^{12}$ Laboratoire de M\'{e}t\'{e}orologie Dynamique/IPSL, Sorbonne Universit\'{e}, ENS, PSL\\ Research University, \'{E}cole Polytechnique, CNRS, Paris, France\\
$^{13}$ Aeolis Research, Chandler, USA \\
$^{14}$ Cavendish Laboratory, University of Cambridge, Cambridge, UK \\

\vspace{8cm}

\noindent \textbf{\textit{Benjamin Benne, Megan A. J. Brown, Bethan S. Gregory, and Paul M. Streeter would like to thank the UK Space Agency for supporting the ACSIOM Workshop and the writing of this white paper through the Exploration Science working group support call.}}

\newpage
\newgeometry{left=2.5cm, right=2.5cm, top=2.5cm, bottom=2.5cm}
\thispagestyle{empty}

\section*{Executive summary}

\quad The ACSIOM workshop was held in Cambridge from 10$^\text{th}$--12$^\text{th}$ November 2025. Its primary aims were twofold: 1) to gather UK researchers involved in Mars atmospheric science to identify current shortcomings and challenges in the field, and 2) to discuss possible collaborations and future steps needed to maximise the research output from current and future Mars missions.
Workshop attendees spanned a wide range of subjects and disciplines, from atmospheric modellers and surface scientists to experimental chemists and Earth/Venus/exoplanet scientists. 
\newline

One of the primary outcomes of the workshop as identified by the attendees was the need for more collaborations between these different communities.
It was agreed that we currently lack sufficient interaction between disciplines and this is a key challenge we need to overcome in order to properly utilise and exploit the vast pool of martian expertise available in the UK. Modellers need reaction rates from experimentalists; experimentalists need modelling context to prioritise their experiments; geologists need modelling context to understand the features they see; modellers need geological context to constrain their results.
Attendees stressed that the UK Mars community has a critical mass which is internationally recognised, and improving the communication bet\-ween communities would allow significant progress to be made in martian research, allowing the UK to maintain leadership in Mars science for years to come.
\newline

Overall, we recommend UKSA support community efforts to start building and stren\-gthening these collaborations.
This could include covering travel costs to allow resear\-chers to meet, visit and use existing laboratory infrastructure (such as experimental faci\-lities to perform reaction rates measurements), and organise in-person workshops, where researchers can discuss and break the jargon barriers which currently impede discussions between modellers, experimentalists, and geologists.
These collaborations should also include researchers whose work is not solely focused on Mars (e.g. Earth, Venus, and exoplanets), to allow for a better return on investment, as our understanding of Mars would greatly benefit from a broader field.
\newline

Finally, we would like to highlight that this is a pivotal moment for such investments as many planetary missions, both for Mars and other terrestrial bodies, are about to be launched or are being considered in the coming years.
Having a strong and close-knit community will be key to better identifying needs and capabilities in order to optimise investments in missions and to have the best scientific return possible.

\newpage

\thispagestyle{empty}
\tableofcontents
\newpage

\setcounter{page}{1}

\section{Introduction}

\quad The Atmospheric Chemistry and Surface Interactions On Mars (ACSIOM) workshop was held in Cambridge between the 10th and 12th of November, 2025. Its objective was to gather UK researchers involved in state-of-the-art research on a variety of Mars atmosphere related topics.
Indeed, the UK Mars community has reached a critical mass and draws on decades of funding and experience, which makes it world-leading in va\-rious topics, such as instrument development, laboratory experiments, data retrievals, and atmospheric modelling.
This workshop therefore gathered 25 researchers, inclu\-ding five keynote speakers and nine Early Career Researchers (ECRs), from different fields, inclu\-ding atmospheric modelling, surface studies, and experimental chemistry, with the aim to spark discussions between the different communities and determine key objectives for future UK Mars science that would help to make significant progress in these various topics.
Keynote speakers were chosen to present the current state of research across the range of topics in order to maximise participation and the breadth of discussions.
\newline

\vspace{-0.2cm}
In this white paper we summarise the contents of the discussion sessions, with the aim of highlighting key steps to be taken within the UK community, and how UKSA could support these initiatives to enable the different research teams to produce cutting-edge science that will allow the UK to remain an international leader in Mars research.
We have divided the paper into four research sections:
\begin{itemize}
    \vspace{-0.1cm}
    \item Atmospheric modelling and the interaction with experimental chemistry
    \vspace{-0.2cm}
    \item Aerosol dust lifting and composition
    \vspace{-0.2cm}
    \item Atmosphere-surface interactions and synthesising models with rover data
    \vspace{-0.2cm}
    \item The utility of comparative planetology
\end{itemize}
We then summarise the content of the paper in Section \ref{sect:summary}.


\begin{figure}[!h]
    \centering
    \includegraphics[width=0.99\linewidth]{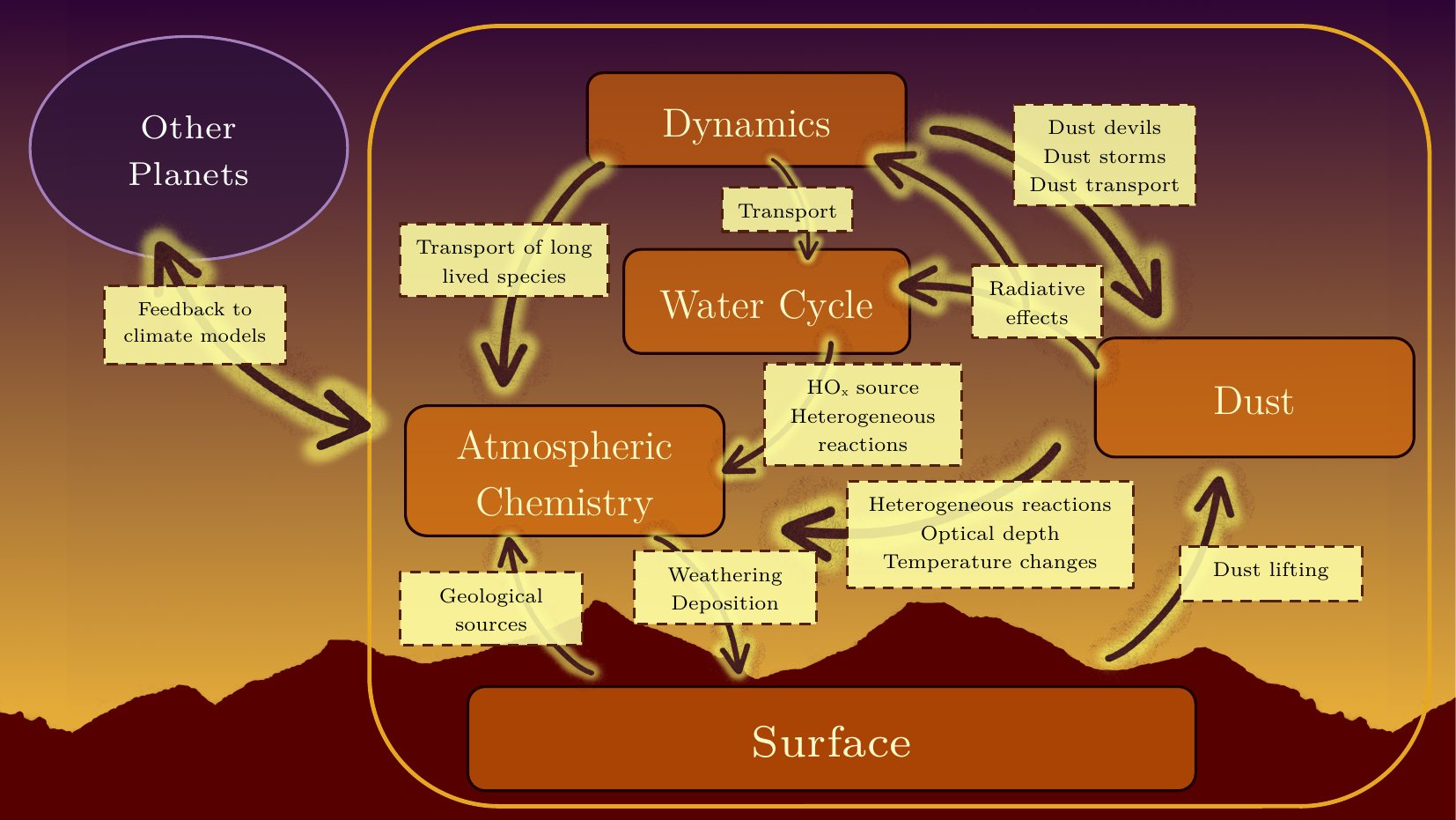}
    \caption{Schematic depicting how each topic interacts with one another and the key processes responsible for this interaction.}
    \label{fig:diagram_topics}
\end{figure}

\newpage

\section{Atmospheric modelling, photochemistry and experimental chemistry}


\quad This section highlights the challenges of experimental chemistry and modelling of the Martian atmosphere, and how these two fields could benefit from collaborations to ensure the UK remains one of the leading countries in Mars atmospheric chemistry.

\subsection{State of the art}

\quad There are currently at least five different Mars General Circulation Models (GCMs) in some form of use in the UK. All these models allow computation of the physics, che\-mistry and dynamics of the Martian atmosphere. The three main Mars GCMs in use in the UK are:
\begin{itemize}
    \item The Mars Planetary Climate Model (MPCM). This model is developed at the La\-boratoire de Météorologie Dynamique (LMD) in Paris, in collaboration with the Atmospheric Space Observations Laboratory (LATMOS, Paris), Oxford University, and the Instituto de Astrofisica de Andalucia (IAA, Granada) \citep{forget_improved_1999,gonzalezgalindo2013,madeleine_revisiting_2011,navarro_global_2014,pottier_unraveling_2017}. This model is used by teams in Edinburgh and Oxford \citep{Gregory2025,benne_3d_2025}.
    \vspace{-0.1cm}
    \item The UK version of the MPCM, the MPCM UK-spectral, is used by teams at the Open University in Milton Keynes \citep{rajendran_global_2025,streeter_global_2025}. This model shares physics and chemistry with the MPCM, but uses a different dynamical core and has data assimilation capabilities (e.g. \citep{Lewis1995,Holmes2020}).
    \vspace{-0.1cm}
    \item The Unified Model (UM) has been developed at the Met Office for the Earth, and is currently being adapted for Mars by teams in Exeter, Bristol, and Cambridge \citep{McCulloch2023}.
\end{itemize}

The MPCM and the MPCM UK-spectral are compatible, sharing the same physical and che\-mical routines, and therefore development of one is beneficial to the other.
The avai\-lability of multiple and independent modelling frameworks is valuable for model intercomparisons and in preventing over-reliance on a single model; similar results from independent different models implies greater confidence in the scientific results being reported.
These tools are critical to better understand the different phenomena influen\-cing the Martian atmosphere by reproducing observed data from different missions, such as, for example, ExoMars Trace Gas Orbiter (TGO), Mars Reconnaissance Orbiter (MRO), or Mars Express. GCMs have significantly improved and grown in complexity over the years, which improved our ability to explain and interpret the increased volume of atmospheric data regarding temperature, aerosols, and chemical species.
\newline

\vspace{-0.2cm}
Modern models are able to reproduce the broad seasonal cycles of dust and water, which significantly influence the Martian atmosphere, but they often lack the fidelity to capture the complex, localised atmospheric features and surface-atmosphere inte\-ractions, such as topography, albedo, thermal inertia, dust cover (e.g. \cite{Viudez-Moreiras2025}). The same situation occurs for trace gases, such as ozone and HCl, for which broad spatial and seasonal variations are correctly reproduced, but direct comparisons with observations are more difficult. The fact that only broad temporal and spatial features can be reproduced indicates that species or processes are missing or not accurately parametrised in models and are a large source of uncertainty \citep{Grenfell2022}.
This includes how we describe atmospheric chemical processes (photolysis, gas-phase or heterogeneous reactions, missing species, volatile cycles) and physical processes (e.g. dust lifting mechanisms, ice aerosol microphysics, electric discharges).
As simulations with these models can be hugely time intensive (individual simulations can take several months), the considered chemical network (i.e. the ensemble of all the chemical reactions) is restricted.
Models only include reactions which are known to occur in the Martian atmosphere, which can lead to gaps (i.e. missing reactions that could, in fact, be very important for current Martian photochemistry).
In addition, critical data used to compute chemical reactions (rate constants, uptake coefficients) are either mis\-sing or predominantly measured under Earth conditions.
Measurements are thus mainly conducted around 300~K (corresponding to room temperature), which is higher than the temperature of the lower Martian atmosphere, which typically varies between 140~K and 250~K.
Some measurements made for the Earth stratosphere cover conditions closer to the Martian atmosphere, but almost none reach temperatures lower than 200~K.
Mo\-dellers therefore use the best data available, but these data are not necessarily suitable for Mars conditions and are often subject to large uncertainties which then propagate into model results.
\newline

\vspace{-0.2cm}
Teams at the Universities of Cambridge, Edinburgh, and Oxford, and the Open University are investigating chemical pathways to try to bridge the gap between models and observations, in particular regarding ozone, which is strongly underestimated by models \citep{brown_role_2024,daerden_heterogeneous_2023,Lefevre2017,lefevre_relationship_2021}. The Open University, Oxford, and Edinburgh teams have also investigated Martian atmospheric HCl, to try to reproduce TGO observations \citep{benne_3d_2025,streeter_global_2025,Gregory2025}. The conclusions of these studies are that he\-terogeneous chemistry strongly influences both ozone and HCl, but uptake coefficients, which are a key parameter to compute the rates of these reactions, are underconstrained, which strongly limits model interpretability to observations.
\newline

\vspace{-0.2cm}
Therefore, it is critical to improve cooperation between modellers and laboratory-based chemists. Modellers can help identify part of the Martian chemical network that need to be revised by the laboratory-based chemists under Martian conditions, e.g. temperature, pressure, high-CO$_2$, appropriate solar irradiance levels. Experimental facilities are already available in the UK:
\begin{itemize}
    \item The Open University has several relevant Mars chambers which are available \footnote{\url{https://university.open.ac.uk/science/physical-science/facilities/hvi-spe-laboratories}}
    \item University of Aberdeen: Mars chamber\footnote{\url{https://www.abdn.ac.uk/geosciences/departments/planetary-sciences/research/facilities/}}: Temperature range = 213-473 K, Pressure range = $<10^{-5}$ mbar to 1 bar, UV solar simulator, relative humidity 0-100\%
    \item University of Leeds: HIRAC (Highly Instrumented Reactor for Atmospheric Che\-mistry)\footnote{\url{https://hirac.leeds.ac.uk/about-hirac/}}: can be used for any bath gas for pressures between $<$ 0.03 mbar (up to $\approx$70$\,$km) to 1 atm, with different lamps to perform photochemical studies, can measure OH and HO$_2$ abundances, temperature range 227-331 K
    \item STFC Diamond Light Source\footnote{\url{https://www.diamond.ac.uk/Instruments/Structures-and-Surfaces/I07/guide/UHV.html}} and Central Laser Facility: could be used to measure surface uptake of gases onto dust and ice aerosols in Martian conditions.
\end{itemize}

To be able to measure reaction rates in conditions spanning the pressure and temperature range of the lowest 70 km of the Martian atmosphere (where HCl and ozone layers are detected), setups should be able to cover pressures from 6 to 10$^{-3}$ mbar, and temperatures from $\approx$ 140 to $\approx$ 250 K.
The UK facilities listed above allow coverage of a significant part of these range, with only small adaptations required to reach the most extreme conditions (very low temperatures in particular).
Therefore, the UK already has many different facilities where key and rare measurements could be performed, which would necessarily allow the UK to become a leader in Mars-related chemical measurements, as these would be used in models around the world.

As an alternative solution, international facilities could be accessible through Europlanet’s Transnational Access (TA) programme. One of the main routes accessible through this call is the Planetary Field Analogues (PFA), which aims at supporting forthcoming missions to Mars, the Moon and Jupiter’s moons. Researchers directly apply to Europlanet to get access to the desired facility, as well as travel and accommodation costs (for up to 2 people). The list of the 19 European facilities available for the 2026 call is available at \url{https://www.europlanet.org/infrastructure/ta-call-2026/}, and include the Open University chambers.
\newline

\vspace{-0.2cm}
In addition to experimental measurements, reaction rates can also be determined with theoretical calculations, but these calculations remain limited despite recent improvements. Indeed, they can determine the correct order of magnitude, but can lead to inaccurate results if the correct “chemical route” is missed. The theoretical technique used to determine the rate constant of a specific reaction also contributes its own uncertainty. It is therefore crucial to continue performing experimental measurements to benchmark theoretical methods, increasing their reliability when experiments are lacking or too difficult to perform.
However, to carry out such measurements or calculations, better communication between the modelling and the experimental chemistry communities is needed, to ensure that measurements are conducted under the right environmental conditions and for key gas-phase and heterogeneous reactions.

\subsection{Challenges}

\quad We list the key challenges below:

\begin{itemize}
    \item We currently lack experimental measurements of rate constants and uptake coefficients covering the entire range of typical Martian conditions. Having these data is crucial to improve calculations in GCMs.

    \item The MPCM UK-spectral and MPCM models currently share a lot of model physics and the same baseline chemical scheme, as they originate from the same original model. This limits our ability to do meaningful cross comparisons between models.


    \item The composition of the Martian dust is not precisely known, nor is the size distribution of dust grains, which is problematic as these two parameters strongly influence heterogeneous reactions. The reactivity of these grains is also unknown, which questions the reliability of rates measured with Earth dust analogues.

    \item There is currently a lack of communication between the chemical modelling and experimental communities, which complicates the identifying of gaps in models, the key reactions which need to be (re)measured, and the knowledge of which facilities are most suitable to perform these measurements.
\end{itemize}

To make significant progress on all these challenges, we have identified specific actions that could be taken directly by the community, that we separated into short-term and long-term actions. We also identified how these actions could be supported by UKSA. These points are listed below.

\subsection{Proposed solutions}

\subsubsection{Short-term actions (2-3 years)}

\paragraph{Modelling community:}
\begin{itemize}
    \item Compile and maintain a prioritised list of chemical reactions that need to be measured/revised to interpret Martian atmospheric photochemistry, based on results from the different modelling teams, to clarify the needs from atmospheric mo\-dellers and experimental chemists.
    \item Continue the development of various models to ensure UK capabilities in Mars modelling, as it allows to highlight uncertainty, strengthen results, point out common flaws, and maintain long-term international collaborations (e.g. between UK and France for the MPCM). In particular, support the development of the UM-Mars model, to allow UK researchers to have access to a model that is significantly different from the PCM family of models. Continue training ECRs to use and improve these models to ensure experience transfer.
    \item Continue collaborations in missions, such as the ongoing ExoMars TGO, or future missions, such as the LightShip-Spotlight or the repurposed Earth Return Orbiter (ERO) from the Mars Sample Return (MSR) mission, to secure access to new measurements to improve and constrain the models (trace gas vertical and temperature profiles, aerosol data). This is also crucial for data assimilation, which relies on observational data, and can help keep/develop instrumental engineering capabilities in the UK.
    \item Create a Slack group to include all the UK modellers to facilitate exchange of information and work updates. Organise annual in-person meetings.
\end{itemize}

\paragraph{Communication between modellers and experimentalists:}
\begin{itemize}
    \item Create a mailing list or a Slack group to foster discussions between the two communities.
    \item Organise regular meetings ($\approx$ twice a year) to present results, discuss possible experiments and collaborations. Organise annual in-person workshops.
    \item Organise visits of experimental facilities, so that modellers can  be aware of what could be measured, and chemists can discuss what needs to be adapted to reach modellers needs.
    \item Make model outputs more widely available and easy to access.
    \item Support and fund PhD students that encourage cross-disciplines of both experimental and modelling chemistry.
\end{itemize}

\paragraph{Experimental chemistry:}
\begin{itemize}
    \item Based on the list of critical chemical reactions to measure, identify facilities where the experiments could be conducted in the specific conditions of the Martian atmosphere (or facilities where the setup could be easily adapted to match these conditions).
    \item Prepare proposals to get access to the facilities and secure funding to perform the experiments. In addition to UK facilities, researchers could identify international facilities accessible through different programs, such as the Europlanet's Transnational Access (TA).
    \item Open these experiments to Earth scientists as well, as Martian conditions can be found in the Earth’s upper atmosphere, as well as other planetary bodies in the solar system. This would also increase the return on every measurement.
\end{itemize}

\subsubsection{Long-term actions ($>$ 3 years)}

\begin{itemize}
    \item Continue applications/experimental campaigns to determine reaction rates in Martian conditions, following the advances made by models.
    \item Create a group to build up an evaluation document that would list relevant rate constants, uptake coefficients, and photochemical parameters of interest for Martian conditions. This group would review the literature to identify the best data for Mars models, to complement JPL evaluations, that are focused on the Earth atmosphere.
    \item Create a common database to store model results from the different teams, as well as kinetics data, so they can be easily accessed and shared. This will allow this data to be used by everyone, therefore increasing the return on investment and making UK science an international benchmark.
    \item Maintaining involvement in space missions, that will allow access to new data to constrain the models and, therefore, point out the needs for new chemical measurements.
    \item As heterogeneous reaction parameters are poorly known, as well as the Martian dust composition, some geoengineering studies on dust analogs could help us understand key parameters influencing dust reactivity, and therefore its potential influence on atmospheric chemistry.
    \item Develop more complex coupled models, including atmosphere, cryosphere, and subsurface processes. These would improve our global understanding of the Martian system, and of the interactions between the different sub-systems.
    \item Perform an inter-comparison between the different UK Mars atmospheric models to evaluate their predictions.
    \item Improve the forecasting capabilities of models, to improve mission planning and prepare ground for human exploration.
    \item Given the amount of data accumulated over the years, the use of machine lear\-ning could also be useful to identify patterns that have not yet been identified by researchers. This could be the influence of some chemical reactions or processes that researchers could then improve in models.
    \item Advocate for future surface instrumentation or experiments using current instrumentation on current surface platforms (i.e. ChemCam/Supercam, MAHLI/WAT\-SON, APXS/PIXL, etc) which can be used to characterise dust, particularly: size, shape, and chemical composition.
\end{itemize}

\subsubsection{How UKSA can support this?}
\begin{itemize}
    \item Continue to invest in missions, such as the ongoing ExoMars TGO, and future missions such as the proposed ESA LightShip/SpotLight, the repurposed ERO, and MEDAL. These investments are crucial to allow the development/exploitation of these missions at ESA, and in turn allows access to high-quality data. Investing in such missions is also beneficial to involve and train ECRs (PhDs and postdocs), who can then contribute to keep scientific expertise in the UK.
    \item Fund PhDs or postdocs projects that would bring the modelling and experimental chemistry communities together.
    \item Fund in-person events (workshops, travel costs for researchers to meet and/or visit facilities) to allow these collaborations to take shape.
    \item Support the building of the Mars' atmosphere photochemical data group, as the outputs provided by this group would be highly valuable at the international level.
    \item Support the purchase of equipment needed to adapt/improve experimental che\-mistry facilities to allow performing measurements in Martian atmospheric conditions.
    \item Create specific calls for driving laboratory data and actually generating data that can be fed into models. Create specific UKSA/STFC call targeting key reaction rate unknowns for Mars models, or a joint UKSA with STFC/NERC call specialising in reaction rates for stratospheric/Mars conditions
    \item If machine learning becomes necessary in some studies, some investments to support teams to invest in appropriate equipment (e.g. GPU servers) would be essential. UKSA could also be involved in existing already existing calls for access to HPC resources (e.g. ARCHER2\footnote{\url{https://www.ukri.org/opportunity/access-to-high-performance-computing-facilities-autumn-2025/}})
\end{itemize}
\newpage

\section{Atmosphere-surface interactions: Dust lifting and dust composition}

\quad Atmospheric dust is a crucial component of the Martian atmospheric system. Dust lifting and transport during dust storms are the main sources of uncertainty and unpredictability in Mars’ weather, notably in the phenomena of martian dust storms which can at irregular intervals (and for unknown reasons) reach global scales. This suspended dust has significant radiative effects via its absorption and scattering of solar and infrared radiation, with corresponding impacts on surface/air temperatures and atmospheric dynamics. In addition, dust has both direct and indirect effects on atmospheric chemistry from indirect impacts on photolysis rates (via atmospheric opacity) to direct involvement in heterogeneous reactions. This is an increasingly active field of research, in which the UK plays a significant role (e.g. \cite{benne_3d_2025,Lewis2007,Mulholland2013,Newman2002,Streeter2020}). Many challenges and open questions remain with great relevance to those interested in Martian surface-atmosphere interactions, atmospheric chemistry, and the meteorological system as a whole.

\subsection{State of the art}

\subsubsection{Dust lifting}

\begin{figure}[!h]
    \centering
    \includegraphics[width=1.0\linewidth]{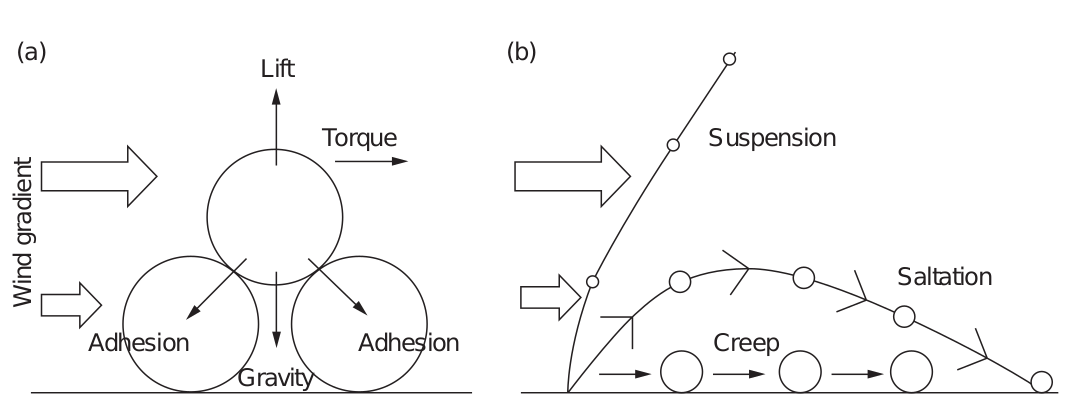}
    \caption{(a) Forces experienced by dust particles and (b) forms of dust particle movement: suspension, saltation, and creep, based on particle mass. Adapted from \cite{Kahre2017}.}
    \label{fig:dust}
\end{figure}

\quad Even now, the fundamentals of dust lifting on Mars are not fully understood. Some dust is known to be lifted by dust devils (convective vortices), which may aid direct lifting and suspension of dust by their stronger convective winds and interior pressure drops. However, the large-scale dust lifting associated with events like dust storms is not linked to dust devils, but wind stress from non-dust devil winds. It is generally thought that dust lifting associated with wind stress occurs by some combination of processes. One is direct lifting: direct suspension of particles by winds. Another process, thought to be key, is saltation: temporary lofting of larger particles which rapidly fall again, their impact causing suspension of smaller particles (see Figure \ref{fig:dust}). These lifting mechanisms have numerous complex relationships to surface factors and the properties of the dust particles themselves, including unknown degrees of interparticle cohesion and potential electrostatic effects. Equations exist and have been used with some success to represent dust lifting on Mars, based on assumptions and semi-empirical relationships derived under Earth conditions.
\newline

\vspace{-0.2cm}

On the one hand, experiments are far easier to conduct with the accessibility of sophisticated equipment which is more difficult to send to Mars. On the other hand, it was questioned how applicable Earth results really are to Martian conditions. Even if Martian temperatures and pressures are represented, we may still be missing important context such as electrostatic effects, the (still unknown) composition of Martian dust, particle size distributions, and more. That said, interest in atmospheric aerosols on Earth has increased in recent years due to their uncertainty in radiative forcing, their role as condensation nuclei, and impact on climate change, all of which may spark future deve\-lopments in sand/dust lifting theory.

\subsubsection{Dust composition}

\quad The composition of Martian dust aerosol is still poorly understood, despite its importance for determining dust optical properties and therefore its radiative impacts. Comparisons of spectral features to known minerals suggest silicates, the result of weathe\-ring of basaltic rocks \citep{Smith2008}, dominate, and dust appears to be enhanced in silicon and chlorine \citep{Berger2016}. The dust population appears to show high global homogeneity \citep{McSween2000}; with analysis suggesting much of it may come from erosion of the same
single sedimentary unit \citep{Ojha2018}. Some work has been done to characterise dust at rover landing sites (e.g. \cite{Berger2016,OConnellCooper2018,Weitz2022}, however these are local characterizations, and may not be representative of global dust compositions.

\subsection{Challenges}

\subsubsection{Dust lifting}

\quad The fundamental triggers of Martian dust lifting—from the cohesive forces between grains to the role of convective vortices—remain largely unresolved. Key questions raised ranged from the highly specific to the general and include:
\begin{itemize}
    \item The need for quantitative measurements (or better estimates) of threshold and cohesion values (specific parameters for dust lifting parametrizations), as part of understanding interparticle forces (see Fig. \ref{fig:dust}.a). Are we adequately representing these forces, or should we be incorporating effects such as coagulation? And how do these affect lifting?
    \item Is saltation or direct lifting dominant in these processes, and in which domains/under what conditions?
    \item What are the roles (if any) of electrostatic and/or vortex suction effects? (e.g. \cite{Wu2026})
    \item What are the most significant differences between Earth and Mars for dust lifting (e.g. planetary boundary layer depth, atmospheric pressure, gravity, aridity)?
    \item How can the above be adequately represented in numerical models?
\end{itemize}

It is generally understood that we are still lacking in our understanding of Martian dust lifting mechanics, with limitations in our ability to perform numerical weather prediction for Mars.

\vspace{-0.3cm}
\subsubsection{Dust composition}

\quad Despite its high homogeneity across the planet, we still do not fully understand the composition of Martian dust. This has implications for understanding of optical properties (and therefore radiative effects) and, in particular, heterogeneous chemical reactions associated with dust. Another key question includes the age of the dust: is it all “old” or is there continual replenishment from the surface? This has implications for weathe\-ring/ageing and the nature of the dust cycle as a whole. It was noted that attempts to characterise possible Martian heterogeneous chemical reactions in Earth laboratories are hampered by the lack of known variables: particle size, composition and shape.

\vspace{-0.2cm}
\subsection{Proposed solutions}

\subsubsection{Dust lifting}

\quad In order to expand our understanding of the physics involved in dust lifting, solutions must come from observations on Mars and laboratory experiments.
It is possible that future progress on this topic depends primarily on close in situ characterisation of Martian dust lifting in progress on Mars. Surface-based measurements to date (such as from the Curiosity and Perseverance rovers and Insight lander (e.g. \cite{Baker2020,Newman2022,Guzewich2023,VincenteRetortillo2023})) have revealed some information about local environmental conditions (e.g. near-surface (1.5\,m) wind speeds, vortex pressure drops, sand availability) associated with raising dust from the Martian surface. However, dedicated surface-based missions or instrumentation to quantitatively measure the properties of lifted dust (vertical flux, particle size distribution) simultaneously with the key environmental variables that drive this lifting (from surface cohesion and sand flux to surface friction velocity and electric field strength) are required for a step-change in our understanding of Martian dust lifting. Such dedicated missions or instruments would have a high impact on our ability to perform weather prediction for Mars, further enabling future exploration and reducing risk to both robotic and human explorers, and are also relevant to astrobiology and planetary protection due to the importance of dust and the circulation on controlling contamination and transport of biological material.

\vspace{-0.2cm}
\subsubsection{Dust composition}

\quad The obvious solution to characterising dust composition is sample return. While expensive and difficult, the advantage of Mars’ apparent global homogeneity is that dust samples from any location will likely be representative of the dust population as a whole. It was also brought up as a question whether laser ablation (e.g. \citep{Ramisetty2018}) could be used as a tool, as for Earth, to characterise aerosol size/composition/mixing ratios. More broadly, the question was raised as to whether existing rover datasets are being under-utilised in studying this question of geochemical characterisation of dust; further discussions are required. One potentially valuable project would be a review paper synthesising all the existing knowledge and literature regarding the chemical composition and optical properties of martian dust; such a paper does not currently exist but would be a boon to modellers, laboratory scientists, and remote sensing observers alike.

\newpage

\section{Rover and lander measurements for surface and atmospheric interactions (specialising in Early Mars)}

\quad Since NASA's Viking landers in the 1970s, Mars rovers and landers (henceforth referred to jointly as rovers) have revealed unprecedented details of the planet’s surface, from organics \citep{Fornaro2025} and silicates \citep{Battersby1997} to ancient lakes \citep{Mangold2021}, and surface winds \citep{Newman2017}. These features (and others) can help us understand Mars' past and present-day climate.

\subsection{State of the art}
\subsubsection{Modern surface observations}

\quad Over the last 30 years, rover missions have been driven largely by the desire to understand ancient surface conditions (e.g. \cite{Crisp2003}) and, more recently, habitability \citep{Grotzinger2012}. In 1997, Pathfinder attempted to measure near surface wind friction speeds and the aerodynamic roughness using the windsock experiment \citep{Sullivan2000}. Although the Mars Exploration Rovers carried no dedicated atmospheric experiments, numerous instruments were employed to observe atmospheric processes.
With the Spirit and Opportunity rovers, the Mini-TES instrument was used to observe atmospheric temperatures, the optical depth of dust, and abundance of water vapour \citep{Smith2006}. Atmospheric opacity observations were taken with Pancam \citep{Lemmon2015}, and dust build up on the solar panels was tracked and used to determine dust settling rates \citep{Stella2010}. Dust devils were tracked by both Spirit in Gusev Crater \citep{Greeley2010} and Opportunity at Meridiani Planum \citep{Murphy2016}, and wheel track erasure rates provided insight into dust settling and sediment saltation rates \citep{Geissler2010}.
\newline

\vspace{-0.2cm}
In 2012, the Mars Science Laboratory (MSL) Rover (Curiosity) landed in Gale Crater, with dedicated instrumentation for measuring environmental variables: the REMS (Rover Environmental Monitoring Station) and RAD (Radiation Assessment Detector) instruments \citep{Grotzinger2012}. REMS was designed to measure six parameters: wind speed, direction and pressure; relative humidity; air and ground temperature, and UV radiation \citep{GomezElvira2014}. Unfortunately, the wind sensor was partially damaged during landing, although the second boom was able to provide wind measurements during the Bagnold dune campaign \citep{Newman2017}. Mastcam and REMS have made systematic measurements of atmospheric aerosols \citep{Campbell2025,Smith2016}, and cloud formation above Gale \citep{Lemmon2024}.
Curiosity's Sample Analysis at Mars (SAM) instrument has measured near-surface atmospheric mixing ratios \citep{Franz2015} including, famously and controversially, methane (e.g. \cite{PlaGarcia2019}.)
\newline

\vspace{-0.2cm}
Most recently, the Insight lander's wind sensors and extremely high-cadence pressure sensors have enabled studies of the local atmosphere including night-time turbulence and transient wave activity (see \cite{Spiga2018} for a review). NASA’s most recent rover, Perseverance, carries the MEDA instrument suite which enables an enhanced set of atmospheric measurements compared to Curiosity \cite{RodriguezManfredi2021,Hueso2023}. In particular, MEDA includes sensors that measure the upward and downward fluxes of thermal and solar (direct and diffuse) radiation, enabling a much better picture of dust content and size distribution within dust clouds and vortices in the rover’s vicinity, as well as the ability to instantaneously detect local dust lifting via albedo changes; these can then be linked to the local meteorology measured at the same time and location \cite{VincenteRetortillo2023}. Furthermore, unlike Curiosity, Perseverance’s wind sensor did not suffer damage on landing, though it did lose functionality later in the mission. A final benefit of Perseverance for studying dust lifting is its location close to Mars’s second most active dust storm track, which resulted in the first observation of local dust lifting activity in an active dust storm location \cite{Lemmon2022}.
\newline

Rovers have provided important insights into local environmental conditions and provided invaluable measurements for validation of global-scale theories of Mars' atmosphere. However, we are naturally constrained by the scarcity of available rovers, the relative clustering of rover locations, and, most relevantly, the remaining unknowns in coupling rover-scale measurements and orbital-scale retrievals/global-scale models.

\subsubsection{Palaeoatmospheric reconstruction}

\quad Understanding the Martian palaeoatmosphere is vital for understanding both present and past habitability. This, however, represents a challenge due to the complexities and difficulties involved in constraining fundamental properties such as pressure and composition. During Mars' early history (before 3.0 Ga) the atmosphere would have played a pivotal role in surface processes: keeping any surface water in a liquid state, permitting formation of the coveted "habitable environments" sought by astrobiologists; facilita\-ting a hydrological cycle, redistributing water across the planets surface; and acting as a significant agent of sediment transport and erosion (for a review of this "warm and wet" Mars see \cite{Wordsworth2016}; for a review of habitability from an astrobiological perspective, see \cite{Fairen2010}.
\newline

\vspace{-0.2cm}
It is established that the Martian atmosphere has largely been lost to space, and mostly before 3.0 Ga \citep{Jakosky2018}; however, little else is strongly established regarding Mars' paleoatmosphere. Atmospheric pressures during this early time period are modelled bet\-ween 0.5--2$\,$ bar \citep{Jakosky2018,Kurokawa2018}, with an upper bound of 4$\,$bar \citep{Kite2014}, and the composition of the primitive atmosphere is largely unknown. The potential presence of wind drag ripples within the stratigraphic record \citep{Lapotre2016} may indicate that, at some point in the past, atmospheric pressure, and therefore sediment transport processes, were similar to that of modern Mars. Indeed, the only physical record of the ancient Martian atmospheric circulation and processes is preserved in the stratigraphic (rock) record; deriving useful data from this comes with its own extensive set of challenges.
\newline

\vspace{-0.2cm}
The stratigraphic record offers an incomplete view of surface processes at best, and any paleoatmospheric signal is largely masked by other -- typically aqueous -- processes. That being said, unique instances of surface conditions related to the atmosphere can be preserved. For example, preserved symmetric ripples \citep{Mondro2025}, formed by wind crossing a shallow body of water, provide a snapshot of the atmosphere at the time. These can tell us that: (1) there was sufficient atmospheric pressure to maintain liquid water at the surface; (2) the body of water was open and ice free, indicating temperatures high enough to prevent freezing; (3) the atmosphere was sufficiently active to form ripples, with the orientation of the ripples indicating prevailing wind direction.\\

\vspace{-0.2cm}
The only reliable records of wind direction are aeolian strata (which are considered undesirable as a habitable environment), which encode wind direction as a proxy of se\-diment transport direction. Studies of ancient aeolian stratigraphy on Mars are rare, and currently, only three separate stratigraphic units have received detailed attention: the Burns formation in Meridiani Planum \citep{Hayes2011}; the Mirador formation in Gale crater \citep{Roberts2025}; and the Stimson formation in Gale crater \citep{Banham2018}. Ignoring the complexities of elucidating trends of wind direction from bulk sediment transport directions \citep{Rubin1987,Rubin2012}, more often than not, only broad trends of wind direction can be identified, and only at the broadest time scales. Seasonal variations in ancient wind direction have been recognised on Mars \citep{Banham2022}; however these are little more than recognition of seasonality in wind direction, with crude indications of wind azimuths. \\

\vspace{-0.2cm}
Data from surface missions have the potential to improve our current understan\-ding of both dynamical and chemical atmospheric-surface interactions, as well as help constrain the uncertainty in the past climate of Mars. The latter is especially important for studying habitability and possible life on ancient Mars (which is also comparable to exoplanets, see Section \ref{sec:planets}).

\subsection{Collaborations}
\subsubsection{Current Collaborations}

\quad A strength of rovers is that they provide a systematic record of modern surface conditions, even when they do not carry a dedicated environmental monitoring package. Opacity observations provide an (almost) daily record of atmospheric optical depth, detailing presence of aerosols (dust and ice). Additional observations relating to dust devils, ripple migration, surface pressure, solar and infrared flux, dust settling rates, and other similar phenomena provide additional information about surface-atmosphere interactions.

A recent study used the SuperCam microphone aboard the Perseverance rover to detect lightning discharges in dust events on Mars \citep{Chide2025}. This discovery has opened a new research field of lightning in Mars’ atmosphere, including the impacts on dust, dynamics, and Martian chemistry. It is a prime example of how creative use of instruments aboard rovers can help us discover more about Mars’ atmosphere.

Rovers can also give us insights into the Martian subsurface in ways which are impossible to achieve from orbital remote sensing, and can elucidate contemporary surface-atmosphere exchange processes. For example, the Phoenix lander famously revealed the presence of subsurface water ice mere centimetres beneath the surface \citep{Mellon2009}, while the Insight lander would have provided the first estimates of surface heat flux if not for its probe underestimating the rockiness of the regolith \citep{Spohn2022}. The planned ESA Ro\-salind Franklin rover will be equipped with a drill to borrow beneath the surface, allowing greater investigation of both current and past surface-atmosphere interactions \citep{DeSanctis2022}.

Atmospheric models can be nested to construct high resolution simulations of areas local to rovers in the present day \citep{Greeley2003}.
This allows for a more direct comparison between models and rovers for geophysical quantities such as surface temperature \citep{Atri2023} and trace gases e.g. ozone \citep{Viudez-Moreiras2025}, which offer further options to evaluate data from both satellite and rover.

\vspace{-0.1cm}
\subsubsection{Potential Collaborations}

\quad One potential use of modelling in surface studies is in estimating the oxidation state of the atmosphere in previous climates from rock types, or extrapolating the seasonality of the climate from ancient lakes by studying the layers of silt and sand \citep{Temel2021}.
Rover data provides the potential for constraining atmospheric pressure at past epochs by using the stratigraphy of rocks. This could help provide some constraint of pressure ranges for ancient Mars, which currently has a large uncertainty range. This kind of analysis requires careful investigation, and allocation of rover resources to collect these data. Such analysis could then be used to feed back into global climate models to help with our understanding of ancient Mars.
\newline

\vspace{-0.2cm}
Another use of modelling is to provide forward models for rock modification/evolu\-tion from ancient Mars conditions to contemporary conditions. From observations, the type of weathering on rocks can be identified based on their types of erosion. Compared to Earth, Mars lacks many destructive modification mechanisms such as tectonic, volcanic, and biological activity; most weathering would be through atmospheric processes (dynamical and chemical). Modellers could provide past climate scenarios for geologists to check against the rock record, in order to further our understanding of erosion in paleoclimates.
\newline

Ripple structures preserved in the stratigraphic record could be used to estimate past atmospheric densities.
Ripples are rarely preserved, but in cases where they are, their morphology may provide information on the pressure conditions in which they occurred. Moreover, their existence is evidence of past squalls or gusts of wind (highly intense winds over short timescales). This may be a topic where Computational Fluid Dynamics (CFD) modelling could provide insights.
\newline

\vspace{-0.2cm}
Climate and CFD models can also be used to provide insights into other rover observations. One example is determining dominant contemporary and/or ancient wind flows based on dune morphology, transverse aeolian ridge (TAR) bedforms, and the stratigraphic record. While degenerate, models can provide ideas of general wind regimes which might be or have been dominant at various times in Mars’ history (e.g. \cite{Barrett2026}. This is an area where limited-domain models can be especially useful for representing smaller-scale, higher-resolution flows associated with specific topographic features. Examples include mesoscale modelling of contemporary aeolian flows \citep{Favaro2025} and CFD modelling of terrestrial dune flows \citep{Jackson2020}.
\newline

\vspace{-0.2cm}
Further collaborations in using rovers as weather stations would be of major benefit to the climate modelling community, particularly as rovers are able to capture geostationary and local weather patterns which are not always possible to record fully with satellites.

\subsection{Challenges}

\quad Below are listed some key challenges for furthering research into martian surface-atmospheric interactions:
\begin{itemize}
    \item Language/jargon barriers for terminology and processes between different fields which limit further collaboration and understanding of the resources and capabi\-lities available. In a similar vein, the lack of communication between the different communities (e.g. atmospheric modellers, surface geologists, geochemists).

    \item Many geological studies focus on understanding aqueous surface environments (lakes and rivers) because of their direct link to habitable surface environments (e.g. \cite{Williams2013}) rather than the atmosphere. 
    While there are some studies that provide information about broad-scale trends of ancient atmospheric-surface interactions, these are very few in number \citep{Hayes2011,Banham2021}.

    \item Rover missions are few and far between, meaning that globally, they provide poor spatial, and temporal resolution for constraining paleoclimate models. This is aggravated by the absence of aeolian deposits at many landing sites - which are the preserved expression of atmospheric-surface interactions.

    \item The lack of awareness of data availability --both from rovers and high-resolution climate simulations-- complicates the communication and collaboration between the surface and modelling communities. There is also a lack of awareness of the preferred data formats used in both disciplines, and how to manipulate them.

    \item There are some gaps in our knowledge of both present day and paleoclimate that still need to be identified. This includes providing geological context for high-resolution/small-scale modelling of martian dynamics, and the inverse of providing modelling support for geological studies of aolian and other landforms.

\end{itemize}





\subsection{Proposed solutions}

\quad Such collaborations between atmospheric modellers and rover scientists are possible with data already available. The key challenge is to connect these two fields together in order to maximise the benefits of both surface data and present/past climate datasets.
\newline

\vspace{-0.2cm}
In order to move forward in utilising data and techniques already available, the lack of communication between atmosphere and surface scientists requires addressing. One first step would be to review the current literature and assess the gaps in the existing atmosphere-surface interaction literature, and instances where there is obvious room for synergy. Secondly, the language barrier between these disciplines could be reduced by either offering more interdisciplinary workshops or supporting more papers which are “translatable” to both subject areas \cite{Andreotti2009}. This latter would help enable the former suggestion.
Another approach to encourage engagement between fields is to make data from rovers more accessible, such as setting up a database of winds and stratigraphy.
\newline

\vspace{-0.2cm}
Supporting the symbiosis between these fields requires little financial investment (e.g. setting up annual workshops to help break down the language barrier between fields, encouraging inter-disciplinary meetings, centralised databases which are accessible to all), and yet allows the community to stay connected and exploit data and tools already available. Further to this, the development of meso- or local-scale modelling would help improve comparability between climate models and rover observations. The areas of modelling context for geology and geological context for modelling would be significantly aided simply by getting the relevant parties to talk in person and discuss exactly what they need in terms of data formats, resolutions, complexity of models, certainty of interpretations, and other such items.
\newpage

\section{Comparative planetology}
\label{sec:planets}

\quad This section discusses the ways in which modelling, observations, and experimental chemistry in support of Mars science can both inform and be informed by ongoing research for other planetary bodies.

\subsection{State of the art}

\textit{Key Mars analogues \& ongoing planetary exploration within and outside of the solar system}
\vspace{2mm}

There are several planetary bodies within the solar system with environments that may provide an illuminating comparison to present-day and early Mars. These include Earth and Venus, as well as Saturn’s largest moon, Titan, and the icy moons Europa, Ganymede, Callisto, and Enceladus. The accelerating discovery of exoplanets is provi\-ding ever-increasing numbers of objects hosting an enormous range of conditions. As the first space-based telescopes capable of characterising the atmospheres of exo-Earths will be deployed in coming decades, the number of known Mars-like or early Mars-like planets will continue to grow. Additionally, our growing insight into early Mars, shar\-pened by modelling and observations, provides something of an analogue for present-day Mars from a different angle. The following provides some examples of ways in which analogue environments can or have been utilised to understand Mars.
\newline

\textit{Earth} has provided the most analogue environments in planetary science, due to re\-latively easy access to a large range of conditions in which human fieldwork and detailed laboratory measurements can be carried out, in addition to remote sensing. For Mars atmospheric science specifically, there is some crossover in relevant conditions. For exam\-ple, Earth’s stratosphere and Mars’ surface have similar temperature and pressure conditions, and incoming radiation results in similar photolysis reaction rates, meaning that extrapolation of chemical reaction rates from Earth to Mars is common. In addition, the potential similarities between early Earth and present day Mars and between early Mars and present day Earth add further opportunities for comparison and offer helpful insights.
\newline

There is also crossover in atmospheric conditions between \textit{Venus} and Mars, which both have CO$_2$-dominated atmospheres. While today they are vastly different worlds, their atmospheric evolution has some shared features, including almost total water loss, as evidenced by ratios of deuterium to hydrogen \citep{Donahue1982,Owen1988}. The lack of a significant water cycle and the resulting effects on trace atmospheric constituents and oxidative che\-mistry, could allow atmospheric chemistry on both pla\-nets to be investigated simultaneously. The Venus clouds have conditions comparable to Earth’s surface, and the aqueous sulfur chemistry on Venus could provide insights into ancient Mars, where surface pressure was greater than in the present and the planet was more volcanically active. Interestingly, HCl is stable in the Venus atmosphere, so a deeper understanding of the chlorine cycle at Venus could shed insight into these species at Mars. As a solar system neighbour which has been visited by several planetary missions over the last few decades and will be targeted by selected missions in the coming years, such as ESA’s EnVision, Venus is a promising analogue to prioritise for comparative planetology.
\newline

Elsewhere in the solar system, \textit{Titan} is slightly smaller than Mars and has a surface pressure of about 1.5 times that of Earth. It hosts stable liquid hydrocarbons on its surface, has an active dust cycle \citep{Rodriguez2018}, and will be explored in much more detail with NASA’s Dragonfly mission, planned for launch in 2028. Titan’s atmospheric composition could be an analogue for a hypothesised transiently-reducing early Mars atmosphere, the composition of which has been proposed as a way in which surface temperatures were kept sufficiently warm to maintain stable liquid water in the past \citep{Wordsworth2017}. \textit{Europa}, \textit{Ganymede}, \textit{Callisto} and \textit{Enceladus}, icy moons of Jupiter and Saturn, join Mars as the top targets in the search for life in the solar system, as they have subsurface liquid water oceans and active geological processes. Europa, Ganymede, and Callisto will be explored in much greater detail in the co\-ming years by NASA’s Europa Clipper and ESA’s Juice mission, while ESA's priority for a future large-class mission is the exploration of Enceladus \citep{Helbert2025}. While much can be learned about early Mars from the Martian geolo\-gical record, stratigraphic observations from time periods of interest can be sparse and difficult to interpret. These moons therefore offer another way to investigate the nature of early Mars.
\newline

While there are a limited number of planets and moons in the solar system, there are a seemingly unlimited number of exoplanets, with more discoveries published each month. Exoplanets therefore constitute a natural planetary laboratory offering a my\-riad of comparison opportunities. Exoplanets can be analogues for Mars (and bodies for which Mars is an analogue) for different reasons. For example, there are rocky exopla\-nets around the same size, temperature, and orbital distance, and planets with thin, Mars-like atmospheres. The James Webb Space Telescope (JWST) can observe rocky planets around cooler M-stars, and the next generation of ground- and space-based telescopes able to characterise the atmospheres of small, rocky planets around brighter stars are being developed. For example, the Habitable Worlds Observatory (HWO) may be able to cha\-racterise the atmospheres of early-Mars-like planets, because Mars-sized planets with thick atmospheres might have sufficiently expansive atmospheres for observation. The Large Interferometer for Exoplanets (LIFE; \cite{Kammerer2022}) is a mission concept designed to characterise the atmospheric composition of a large number of small, rocky exoplanets orbiting a range of host stars. The Extremely Large Telescope (ELT), currently under development, will be able to detect rocky planets in the habitable zones of stellar systems (e.g., \cite{Padovani2023}. As well as details of atmospheric composition, future measurements may be able to capture short-term temporal variability. For example, dynamic silicate clouds have already been detected in some hot, planetary-mass objects (e.g., VHS 1256 b; \cite{Miles2023}. An intriguing possibility for observations by the next generation of telescopes could be the detection of mineral dust aerosol storms comparable to those on Mars. Comparison of rocky exoplanets to Mars will help us speculate about the processes operating elsewhere in the galaxy, and inform us of the importance of factors controlling planetary atmospheres and habitability.

\subsection{Opportunities and Challenges}

\quad Inter-planetary comparison offers a number of avenues for continued discovery of early and present-day Mars, and applications of Mars science to other systems. In the context of atmospheric chemistry and surface interactions, about which our workshop particularly focused, we highlighted three example topic areas for which planetary comparison provides exciting opportunities.

\begin{itemize}
    \item \textit{Answers to questions in atmospheric chemistry}

    What can reaction rates measured on Earth (in Earth-like or artificial conditions) reveal about likely reaction rates in a range of extraterrestrial environments with, e.g., different pressures, temperatures and atmospheric compositions? What processes, sources, and sinks control the distribution and concentrations of atmospheric constituents, but are missing in atmospheric models?

    \item \textit{Understanding of atmospheric processes (e.g., the dust cycle)}

    What can terrestrial dust storms (or dust storms further afield) tell us about dust activity in Mars’ atmosphere? What natural or artificial material would make a good analogue for the composition of dust particles—in terms of composition, size, and shape—and what might laboratory experiments on such particles reveal?

    \item \textit{Insights into how planetary atmospheres evolve}

    What processes shape planetary atmospheres, and how do these factors vary bet\-ween different objects? What was Mars’ early climate like? Was Mars ever habi\-table?
\end{itemize}

We also identified three challenges faced by a planetary comparison approach. Firstly, processes do not necessarily operate in the same way in the vastly different environments represented by planetary bodies. For example, Mars rovers have observed geological bedforms that are not seen on Earth due to the much lower atmospheric pressure \citep{Lorenz2014}. The heterogeneous reaction of HO$_2$ on the surface of ice particles is considered important for understanding the distribution and seasonality of trace gases in the Martian atmosphere, but there is only one data point measured for Earth conditions \citep{Cooper1996}, and further measurements are urgently needed. Heterogeneous reactions occurring on dust particles are also important, but measurements made on Earth are not necessa\-rily representative because of the much greater humidity, as well as uncertainty in Mars dust composition and age. Terrestrial dust activity can be used as an analogue for Martian dust events, but it occurs on a much smaller scale than planet-encircling global dust storms that happen on the Red Planet. Further, different engineering challenges for missions to different planetary systems mean that direct comparison can be limited.
\newline

\vspace{-0.2cm}
Secondly, planetary comparison relies, to a significant extent, on data from missions, which necessarily comes with limitations. For example, new planetary mission concepts are competitive and expensive, with considerable uncertainties and risks. Even when a mission concept is selected, developed, and launched, the reality does not preclude long wait times before the instruments reach their destination. Further, planetary missions are technologically restricted and both measurements and data collection rates are limi\-ted.
\newline

\vspace{-0.2cm}
Thirdly, it was observed that scientific communities can often centre on a specific planetary body or target-specific mission, and that communication and collaboration channels between such communities could be utilised more effectively to facilitate comparative planetology. This is successfully achieved in some interdisciplinary fields, such as astrobiology and biosignature science, but is not universal.

\subsection{Proposed solutions}

\quad In light of these opportunities and challenges, we recommend actions to enhance collaboration between research communities, which will address all three challenges above. The ACSIOM Workshop was a successful example of this on a small scale. Workshop de\-legates included experts on Mars, Earth, Venus and exoplanets; a session was dedicated to Comparative Planetology; and participants had ample opportunities to engage with others from different fields during discussion sessions and breaks, in which cross-disciplinary conversations were encouraged. As a result, it was frequently commented that the breadth of the programme, the cross-disciplinarity and the interplanetary discussion were highlights of the workshop. Further, when asked who should be invited if a similar workshop was run again, some suggested that more researchers working on Earth and other planets would be valuable.
Therefore, we suggest that opportunities to forge connections between different communities at existing conferences are leve\-raged further. Some potential examples include taking care in scheduling concurrent oral sessions, cross-disciplinary social events, and steps to break down scientific language barriers between different groups. From the particular sub-field of understan\-ding Mars atmosphere-surface interactions, we also propose that conducting a follow-up ACSIOM-like workshop would be useful for best integrating ongoing work within the UK planetary science community.
We expect that the resulting increase in interplanetary research discussion will have the following effects:
\begin{itemize}
\vspace{-0.1cm}
    \item \textit{Enhanced planetary mission and science effectiveness}. Knowledge exchange will increase the potential for new missions and mission concepts to be informed by previous missions to other planetary bodies. It will also allow for results discovered by one group to be utilised or adapted for other applications.
    \item \textit{Enhanced planetary mission and science engagement and impact}. Knowledge exchange will also increase input from the wider planetary community on certain planetary missions and projects, regarding e.g., mission goals and design, which will increase impact and cross-community engagement.
    \vspace{-0.1cm}
    \item \textit{New interplanetary collaborations and projects}. This could include the design of new experiments, models or instruments for multiple targets.
    \vspace{-0.1cm}
    \item \textit{Application of cutting-edge Earth studies to planetary science}. As novel tools, techniques, measurements, and modelling applied to the Earth develop, knowledge exchange will allow such sophisticated methods to be applied to other worlds.

\end{itemize}

\vspace{+0.1cm}
One outcome of these discussions could be the construction a document which sets out a list of “community” priorities for future missions. This could be further broken down into levels of priority (how important is this to the community?) and scope (is this a mission-level concept, or some useful science which could be done with an instrument which has a different primary focus?). While not meant to be in any sense binding, this could serve as a useful barometer of priorities and opinions, and a repository of both big ideas (e.g. mission-level concepts such as ice-core drilling) and spin-off science (e.g. radio occultations for communication can also provide temperature profiles.
With these priorities in mind, the community could then review different mission concepts and start being involved in missions that are best aligned with these priorities. This would allow to concentrate the effort on a smaller number of missions and improve the collaboration between researchers and instrumental teams, while increasing the output for various research communities.

\newpage

\section{Summary}
\label{sect:summary}

\quad The ACSIOM workshop was a great success in bringing interdisciplinary fields together, from surface geology of Mars to atmospheric chemistry.
Attendees enjoyed the keynote talks, and also the dedicated discussion sessions with colleagues from across different disciplines and planets. The key takeaways from this workshop are:

\begin{itemize}
    \item We need \textbf{more opportunities for workshops} to help further collaborations \textbf{bet\-ween modellers, experimentalists, and surface geologists}; and between Mars scientists and Earth/exoplanet scientists.
    \item A \textbf{better communication between lab-based experimentalists and modellers} must allow to better identify key measurements to be undertaken and identify facilities where these measurements could be performed in relevant Martian conditions.
    \item Surface scientists and atmospheric scientists currently \textbf{lack a good enough understanding of each others' work and capabilities}, and how they might help each other; this is partly due to a lack of common language and partly due to a lack of regular interactions.
    \item While foundational efforts such as \textbf{model development, reaction rate characterization, and dataset curation} lack the immediate appeal of high-impact discoveries, they \textbf{constitute the critical infrastructure} upon which all future Martian science is built.
    \item \textbf{Support of Mars science and exploration missions} is crucial for \textbf{maintaining UK leadership in this field}, and in particular the proposed ESA SpotLight/LightShip programme which will assert UK dominance in Mars atmospheric and surface science.
\end{itemize}

In addition, this workshop demonstrated that the UK has a critical mass of Mars researchers that are keen to start more cross-disciplinary collaborations to utilise the available data and tools, as well as strengthen the impact of UK science on international scale.

To this day, four Mars missions from four different space agencies are planned or en-route: ESA's Rosalind Franklin rover, JAXA's MMX sample return from Mars' satellites, CNSA's Tianwen-3  return sample mission, and NASA's ESCAPADE for upper atmosphere processes.
In addition, a lot of Mars mission concepts are already being studied: ESA's orbiters (LightShip/Spotlight, repurposed ERO, M-MATISSE), orbiter constellation MACAWS, and landers (MEDAL), and NASA/JAXA/CSA/ASI i-MIM orbiter.
Further to this, there are missions targeting other planetary bodies which would be beneficial for Mars science, such as Venus (ESA's EnVision), the Jupiter system (ESA's JUICE, NASA's Europa Clipper), Titan (NASA's Dragonfly), and exoplanets (NASA's Habitable World Observatory, ESA's LIFE interferometer).

Despite the complex international context and the many cuts in research budgets, this list of set missions promises new and exciting high-quality planetary data, which can be used either directly or indirectly for improving our understanding of Mars.
Hence, this era is pivotal to encourage and support community-led initiatives, as outlined in this white paper, to maintain the UK at the forefront of Mars research for years to come.

\newpage

\printbibliography

@article{forget_improved_1999,
	title = {Improved general circulation models of the {Martian} atmosphere from the surface to above 80 km},
	volume = {104},
	copyright = {http://onlinelibrary.wiley.com/termsAndConditions\#vor},
	issn = {0148-0227},
%	url = {https://agupubs.onlinelibrary.wiley.com/doi/10.1029/1999JE001025},
	doi = {10.1029/1999JE001025},
	abstract = {We describea setof two "new generation"generalcirculationmodelsof the Martian atmospherederivedfrom themodelswe originallydevelopedin the early 1990s. The twonewmodelssharethesamephysicalparameterizationbsutusetwocomplementary numericalmethodsto solvethe atmosphericdynamicequations.The verticalresolution near the surface has been refined, and the vertical domain has been extended to above 80 km. Thesechangesare accompaniedby the inclusionof state-of-the-arpt arameterizationtso bettersimulatethe dynamicalandphysicalprocessesnero'the surface(boundarylayer schemes, ubgrid-scalteopographyparameterizatione,tc.) andathighaltitude(gravitywave drag).In addition,radiativetransfercalculationsandtherepresentatioonf polarprocesses havebeensignificantlyimproved.We presentsomeexamplesof zonal-meanfieldsfrom simulationsusingthe modelat severalseasons.One relativelynovelaspect,previously introducedby Wilson[1997], is that aroundnorthernwinter solsticethe strongpole to polediabaticforcingcreatesa quasi-global,angular-momentumconservingHadley cell whichhasno terrestrialequivalent.Within sucha cell the Coriolisforcesacceleratethe wintermeridionalflow towardthepole andinducea strongwarmingof the middlepolar atmospherdeownto 25 km. Thiswinterpolarwarminghadbeenobservedbutnotproperly modeleduntil recently.In fact, thermalinversionsaregenerallypredictedaboveone,and oftenboth,polesm'ound60-70 km. However,theMars middleatmosphereabove40 km is foundto bevery model-sensitiveandthusdifficultto simulateaccuratelyin the absenceof observations.},
	number = {E10},
%	urldate = {2024-07-02},
	journal = {J. Geophys. Res. Planets},
	author = {Forget, François and Hourdin, Frédéric and Fournier, Richard and Hourdin, Christophe and Talagrand, Olivier and Collins, Matthew and Lewis, Stephen R. and Read, Peter L. and Huot, Jean‐Paul},
	year = {1999},
	pages = {24155--24175},
	file = {Forget et al. - 1999 - Improved general circulation models of the Martian.pdf:C\:\\Users\\bbenne\\Zotero\\storage\\SFFVKSMT\\Forget et al. - 1999 - Improved general circulation models of the Martian.pdf:application/pdf},
}

@article{daerden_heterogeneous_2023,
	title = {Heterogeneous {Processes} in the {Atmosphere} of {Mars} and {Impact} on {H} $_{\textrm{2}}$ {O} $_{\textrm{2}}$ and {O} $_{\textrm{3}}$ {Abundances}},
	volume = {128},
	issn = {2169-9097, 2169-9100},
%	url = {https://agupubs.onlinelibrary.wiley.com/doi/10.1029/2023JE008014},
	doi = {10.1029/2023JE008014},
	abstract = {Current models underestimate the highest observed ozone (O3) column densities on Mars. These estimates could be improved by including the uptake of odd hydrogen species (HOx) on water ice clouds, but the reported uptake coefficient of HO2 is likely overestimated for atmospheric conditions. This leaves a fundamental problem in Mars' atmospheric chemistry unsolved. Here, using the GEM-Mars general circulation model, we explore a range of processes involving multiple phases (gas, adsorbed and solid) that may contribute to an alternative solution. First, we focus on hydrogen peroxide (H2O2) and discuss its physical states on Mars and its chemical impact. We also conjecture its photolytic destruction in ices with model simulations and Compact Reconnaissance Imaging Spectrometer for Mars observations. Then, we include in the model all relevant (for Mars) heterogeneous reactions, both on dust and water ice, recommended by the International Union of Pure and Applied Chemistry for terrestrial atmospheric studies. We find that only the uptake of HO2 and H2O2 on dust are efficient on Mars. Finally, we find that attenuation of sunlight by water ice clouds in the calculation of photolysis rates leads to increased O3 and H2O2 abundances below the ice clouds. The combination of the proposed processes leads to O3 increases without the need for strong uptake of HO2 on ice, but it remains difficult to find a good agreement with O3 and H2O2 observations on the global scale. We provide specific recommendations for future work in observations, laboratory experiments and modeling to advance our understanding of fundamental chemistry on Mars.},
	language = {en},
	number = {12},
%	urldate = {2024-07-02},
	journal = {Journal of Geophysical Research: Planets},
	author = {Daerden, Frank and Crowley, John N. and Neary, Lori and Smith, Michael D. and Loeffler, Mark J. and Clancy, R. Todd and Wolff, Michael J. and Aoki, Shohei and Sagawa, Hideo},
	month = dec,
	year = {2023},
	pages = {e2023JE008014},
	file = {Daerden et al. - 2023 - Heterogeneous Processes in the Atmosphere of Mars .pdf:C\:\\Users\\bbenne\\Zotero\\storage\\S6B6569R\\Daerden et al. - 2023 - Heterogeneous Processes in the Atmosphere of Mars .pdf:application/pdf},
}

@article{lefevre_relationship_2021,
	title = {Relationship {Between} the {Ozone} and {Water} {Vapor} {Columns} on {Mars} as {Observed} by {SPICAM} and {Calculated} by a {Global} {Climate} {Model}},
	volume = {126},
	issn = {2169-9097, 2169-9100},
%	url = {https://agupubs.onlinelibrary.wiley.com/doi/10.1029/2021JE006838},
	doi = {10.1029/2021JE006838},
	abstract = {Ozone (O3) in the atmosphere of Mars is produced following the photolysis of CO2 and is readily destroyed by the hydrogen radicals (HOx) released by the photolysis and oxidation of water vapor. As a result, an anti-correlation between ozone and water vapor is expected. We describe here the O3-H2O relationship derived from 4 Martian years of simultaneous observations by the SPICAM spectrometer onboard the Mars Express spacecraft. A distinct anti-correlation is found at high latitudes, where the O3 column varies roughly with the −0.6 power of the H2O column. The O3 and H2O columns are uncorrelated at low latitudes. To evaluate our quantitative understanding of the Martian photochemistry, the observed O3-H2O relationship is then compared to that predicted by a global climate model with photochemistry. For identical model and observed abundances of H2O, the model underpredicts observed ozone by about a factor of 2 relative to SPICAM when using the currently recommended gas-phase chemistry. Sensitivity studies employing low-temperature CO2 absorption cross sections, or adjusted kinetics rates, do not solve this bias. Taking into account potential heterogeneous processes of HOx loss on clouds leads to a significant improvement, but only at high northern latitudes. More broadly, the modeled ozone deficits suggest that the HOx-catalyzed photochemistry is too efficient in our simulations. This problem is consistent with the long-standing underestimation of CO in Mars photochemical models, and may be related to similar difficulties in modeling O3 and HOx in the Earth's upper stratosphere and mesosphere.},
	language = {en},
	number = {4},
%	urldate = {2024-07-02},
	journal = {J. Geophys. Res. Planets},
	author = {Lefèvre, F. and Trokhimovskiy, A. and Fedorova, A. and Baggio, L. and Lacombe, G. and Määttänen, A. and Bertaux, J.‐L. and Forget, F. and Millour, E. and Venot, O. and Bénilan, Y. and Korablev, O. and Montmessin, F.},
	month = apr,
	year = {2021},
	pages = {e2021JE006838},
	file = {Lefèvre et al. - 2021 - Relationship Between the Ozone and Water Vapor Col.pdf:C\:\\Users\\bbenne\\Zotero\\storage\\AYAA9PHA\\Lefèvre et al. - 2021 - Relationship Between the Ozone and Water Vapor Col.pdf:application/pdf},
}

@article{navarro_global_2014,
	title = {Global climate modeling of the {Martian} water cycle with improved microphysics and radiatively active water ice clouds},
	volume = {119},
	copyright = {©2014. American Geophysical Union. All Rights Reserved.},
	issn = {2169-9100},
%	url = {https://onlinelibrary.wiley.com/doi/abs/10.1002/2013JE004550},
	doi = {10.1002/2013JE004550},
	abstract = {AbstractWater ice clouds play a key role in the radiative transfer of the Martian atmosphere, impacting its thermal structure, its circulation, and, in turn, the water cycle. Recent studies including the radiative effects of clouds in global climate models (GCMs) have found that the corresponding feedbacks amplify the model defaults. In particular, it prevents models with simple microphysics from reproducing even the basic characteristics of the water cycle. Within that context, we propose a new implementation of the water cycle in GCMs, including a detailed cloud microphysics taking into account nucleation on dust particles, ice particle growth, and scavenging of dust particles due to the condensation of ice. We implement these new methods in the Laboratoire de Météorologie Dynamique GCM and find satisfying agreement with the Thermal Emission Spectrometer observations of both water vapor and cloud opacities, with a significant improvement when compared to GCMs taking into account radiative effects of water ice clouds without this implementation. However, a lack of water vapor in the tropics after Ls = 180° is persistent in simulations compared to observations, as a consequence of aphelion cloud radiative effects strengthening the Hadley cell. Our improvements also allow us to explore questions raised by recent observations of the Martian atmosphere. Supersaturation above the hygropause is predicted in line with Spectroscopy for Investigation of Characteristics of the Atmosphere of Mars observations. The model also suggests for the first time that the scavenging of dust by water ice clouds alone fails to fully account for the detached dust layers observed by the Mars Climate Sounder.},
	language = {en},
	number = {7},
%	urldate = {2024-11-21},
	journal = {J. Geophys. Res. Planets},
	author = {Navarro, T. and Madeleine, J.-B and Forget, F. and Spiga, A. and Millour, E. and Montmessin, F. and Määttänen, A.},
	year = {2014},
	note = {\_eprint: https://agupubs.onlinelibrary.wiley.com/doi/pdf/10.1002/2013JE004550},
	keywords = {Mars, atmosphere, climate, clouds, global climate model, water},
	pages = {1479--1495},
	file = {Full Text PDF:C\:\\Users\\bbenne\\Zotero\\storage\\KP429MSD\\Navarro et al. - 2014 - Global climate modeling of the Martian water cycle.pdf:application/pdf;Snapshot:C\:\\Users\\bbenne\\Zotero\\storage\\RMPRBBQL\\2013JE004550.html:text/html},
}

@article{pottier_unraveling_2017,
	title = {Unraveling the martian water cycle with high-resolution global climate simulations},
	volume = {291},
	issn = {0019-1035},
%	url = {https://www.sciencedirect.com/science/article/pii/S0019103516301026},
	doi = {10.1016/j.icarus.2017.02.016},
	abstract = {Global climate modeling of the Mars water cycle is usually performed at relatively coarse resolution (200−300km), which may not be sufficient to properly represent the impact of waves, fronts, topography effects on the detailed structure of clouds and surface ice deposits. Here, we present new numerical simulations of the annual water cycle performed at a resolution of 1° × 1° (∼ 60 km in latitude). The model includes the radiative effects of clouds, whose influence on the thermal structure and atmospheric dynamics is significant, thus we also examine simulations with inactive clouds to distinguish the direct impact of resolution on circulation and winds from the indirect impact of resolution via water ice clouds. To first order, we find that the high resolution does not dramatically change the behavior of the system, and that simulations performed at ∼ 200 km resolution capture well the behavior of the simulated water cycle and Mars climate. Nevertheless, a detailed comparison between high and low resolution simulations, with reference to observations, reveal several significant changes that impact our understanding of the water cycle active today on Mars. The key northern cap edge dynamics are affected by an increase in baroclinic wave strength, with a complication of northern summer dynamics. South polar frost deposition is modified, with a westward longitudinal shift, since southern dynamics are also influenced. Baroclinic wave mode transitions are observed. New transient phenomena appear, like spiral and streak clouds, already documented in the observations. Atmospheric circulation cells in the polar region exhibit a large variability and are fine structured, with slope winds. Most modeled phenomena affected by high resolution give a picture of a more turbulent planet, inducing further variability. This is challenging for long-period climate studies.},
%	urldate = {2024-11-21},
	journal = {Icarus},
	author = {Pottier, Alizée and Forget, François and Montmessin, Franck and Navarro, Thomas and Spiga, Aymeric and Millour, Ehouarn and Szantai, André and Madeleine, Jean-Baptiste},
%	month = jul,
	year = {2017},
	keywords = {Mars, Atmospheres dynamics, Climate, Mars atmosphere, Meteorology},
	pages = {82--106},
	file = {Submitted Version:C\:\\Users\\bbenne\\Zotero\\storage\\FRYAAIHT\\Pottier et al. - 2017 - Unraveling the martian water cycle with high-resol.pdf:application/pdf},
}

@article{madeleine_revisiting_2011,
	title = {Revisiting the radiative impact of dust on {Mars} using the {LMD} {Global} {Climate} {Model}},
	volume = {116},
	issn = {0148-0227},
%	url = {https://agupubs.onlinelibrary.wiley.com/doi/full/10.1029/2011JE003855},
	doi = {10.1029/2011JE003855},
	abstract = {Airborne dust is the main driver of Martian atmospheric temperature, and accurately accounting for its radiative effect in Global Climate Models (GCMs) is essential. This requires the modeling of the dust distribution and radiative properties, and when trying to simulate the true climate variability, the use of the observed dust column opacity to guide the model. A recurrent problem has been the inability of Mars GCMs to predict realistic temperatures while using both the observed dust radiative properties and column opacity. One would have to drive the model with a tuned opacity to reach an agreement with the observations, thereby losing its self-consistency. In this paper, we show that using the most recently derived dust radiative properties in the LMD (Laboratoire de Météorologie Dynamique) GCM solves this problem, which was mainly due to the underestimation of the dust single scattering albedo in the solar domain. However, an overall warm temperature bias remains above the 1 hPa pressure level. We therefore refine the model by implementing a ?semi-interactive? dust transport scheme which is coupled to the radiative transfer calculations. This scheme allows a better representation of the dust layer depth in the model and thereby removes the remaining warm bias. The LMD/GCM is now able to predict accurate temperatures without any tuning of the dust opacity used to guide the model. Remaining discrepancies are discussed, and seem to be primarily due to the neglect of the radiative effect of water-ice clouds, and secondarily to persisting uncertainties in the dust spatial distribution.},
	number = {E11},
%	urldate = {2024-11-21},
	journal = {J. Geophys. Res. Planets},
	author = {Madeleine, J.-B. and Forget, F. and Millour, E. and Montabone, L. and Wolff, M. J.},
%	month = nov,
	year = {2011},
	note = {Publisher: John Wiley \& Sons, Ltd},
	keywords = {Mars, aerosols, climate, dust, modeling, scattering},
	file = {Full Text:C\:\\Users\\bbenne\\Zotero\\storage\\NIMKY437\\Madeleine et al. - 2011 - Revisiting the radiative impact of dust on Mars us.pdf:application/pdf},
}

@article{rajendran_global_2025,
	title = {Global {Transport} of {Chlorine} {Species} in the {Martian} {Atmosphere} and the {Resulting} {Surface} {Distribution} of {Perchlorates}},
	volume = {130},
	copyright = {© 2025. The Author(s).},
	issn = {2169-9100},
%	url = {https://onlinelibrary.wiley.com/doi/abs/10.1029/2024JE008537},
	doi = {10.1029/2024JE008537},
	abstract = {Recent observations by instruments aboard the ExoMars Trace Gas Orbiter (TGO) have revealed the seasonal presence of hydrogen chloride (HCl {\textbackslash}textHCl{\textbackslash}) in the Martian atmosphere. This discovery may have important implications for Martian photochemistry as chlorine species are chemically active, and it may provide a link between the atmosphere and known surface reservoirs of chlorine. However, the global distribution of atmospheric HCl {\textbackslash}textHCl{\textbackslash} is unknown beyond the very sparse TGO observations, and the source and sink processes driving the observed variability of HCl {\textbackslash}textHCl{\textbackslash} are not currently understood. We used a Martian global climate model to investigate, for the first time, the spatial distribution of chlorine species in the Martian atmosphere, and the resulting distribution of surface perchlorates formed via adsorption of atmospheric chlorine species. We adapted an existing Martian photochemical scheme to include gas-phase chlorine chemistry with HCl as the source species, and the resulting atmospheric perchloric acid was allowed to deposit onto the Martian surface via a heterogeneous adsorption scheme. We found that odd-oxygen (O,O3 {\textbackslash}mathrmO,{\textbackslash}mathrmO\_3{\textbackslash}) and odd-hydrogen (H,OH,HO2 {\textbackslash}mathrmH,{\textbackslash}textOH,{\textbackslash}textHO\_2{\textbackslash}) species play a major role in controlling the distribution of atmospheric chorine species. Surface perchlorate deposition was found to occur preferentially at high latitudes; in the tropics, the perchlorate distribution was anti-correlated with surface thermal inertia and agreed qualitatively with observations of surface chlorine. Our model predicted a relative enhancement of HCl in polar regions, but it did not reproduce the observed strong seasonality of HCl, suggesting that heterogeneous chemistry may be required to explain the observed chlorine cycle.},
	language = {en},
	number = {3},
%	urldate = {2025-04-30},
	journal = {J. Geophys. Res. Planets},
	author = {Rajendran, K. and Streeter, P. M. and Lewis, S. R. and Duffy, M. K. D. and Holmes, J. A. and Olsen, K. S. and Korablev, O. and Patel, M. R.},
	year = {2025},
	note = {\_eprint: https://agupubs.onlinelibrary.wiley.com/doi/pdf/10.1029/2024JE008537},
	keywords = {perchlorates, atmospheric chemistry, Mars, chlorine, atmospheric modeling},
	pages = {e2024JE008537},
	file = {Full Text PDF:C\:\\Users\\bbenne\\Zotero\\storage\\CZIYMY9V\\Rajendran et al. - 2025 - Global Transport of Chlorine Species in the Martia.pdf:application/pdf;Snapshot:C\:\\Users\\bbenne\\Zotero\\storage\\9HRPXYXA\\2024JE008537.html:text/html},
}

@article{streeter_global_2025,
	title = {Global {Distribution} and {Seasonality} of {Martian} {Atmospheric} {HCl} {Explained} {Through} {Heterogeneous} {Chemistry}},
	volume = {52},
	copyright = {© 2025. The Author(s).},
	issn = {1944-8007},
%	url = {https://onlinelibrary.wiley.com/doi/abs/10.1029/2024GL111059},
	doi = {10.1029/2024GL111059},
	abstract = {Recent observations from the ExoMars Trace Gas Orbiter (TGO) have revealed the presence of hydrogen chloride (HCl) in the martian atmosphere. HCl shows strong seasonality, primarily appearing during Mars' perihelion period before decreasing faster than projected from photolysis and gas-phase chemistry. HCl profiles also display local anti-correlation with water ice aerosol. One candidate explanation is heterogeneous chemistry. We present the first results from a heterogeneous chlorine chemistry scheme incorporated into a Mars global climate model (GCM), with atmospheric dust/water ice parameterized as an HCl source/sink respectively. Results were compared against a Mars GCM with gas-phase only chlorine chemistry and observations from TGO's Atmospheric Chemistry Suite. We found that the heterogeneous scheme significantly improved the modeled HCl seasonal, latitudinal, and vertical distribution, supporting a crucial role for heterogeneous chemistry in Mars' chlorine cycle. Remaining discrepancies show that further work is needed to characterize the exact aerosol reactions involved.},
	language = {en},
	number = {6},
%	urldate = {2025-06-12},
	journal = {Geophysical Research Letters},
	author = {Streeter, Paul M. and Rajendran, Kylash and Lewis, Stephen R. and Olsen, Kevin S. and Trokhimovskiy, Alexander and Korablev, Oleg and Patel, Manish R.},
	year = {2025},
	note = {\_eprint: https://agupubs.onlinelibrary.wiley.com/doi/pdf/10.1029/2024GL111059},
	keywords = {aerosols, Mars atmosphere, chlorine, GCM, heterogeneous chemistry},
	pages = {e2024GL111059},
	file = {Full Text PDF:C\:\\Users\\bbenne\\Zotero\\storage\\ILYMRLG5\\Streeter et al. - 2025 - Global Distribution and Seasonality of Martian Atm.pdf:application/pdf;Snapshot:C\:\\Users\\bbenne\\Zotero\\storage\\U8KGSZZW\\2024GL111059.html:text/html},
}

@article{brown_role_2024,
	title = {The {Role} and {Lifetime} of {Dissociative} {Heterogeneous} {Processes} in {Improving} {Simulated} {Ozone} on {Mars}},
	volume = {129},
	issn = {2169-9097, 2169-9100},
%	url = {https://agupubs.onlinelibrary.wiley.com/doi/10.1029/2024JE008302},
	doi = {10.1029/2024JE008302},
	abstract = {Ozone simulated in Mars Global Climate Models (MGCMs) is used to assess the underlying chemistry occurring in the atmosphere. Currently, ozone total column abundance (TCA) is under‐predicted in MGCMs by up to 120\%, implying missing or inaccurate chemistry in models. Heterogeneous reactions of hydroxyl radicals (HOX) have been offered as an explanation for some of this bias, because they cause ozone to increase at locations where it's currently under‐predicted. We use four simulations to compare modeled ozone TCA with observations from the UVIS spectrometer aboard the ExoMars Trace Gas Orbiter to improve the representation of heterogeneous processes and their impact on ozone. We use a gas‐phase only run, a dissociative scheme, an adsorbed HOX retention scheme, and a hybrid scheme that combines the dissociative mechanism with the retention of HOX on water ice. We find retention of HOX is dependent on water ice sublimation, and ozone abundance increases when water ice persists for longer periods (1–20 sols). Over time, the loss of HOX causes a depletion in H2O2 concentration (HOX reservoir), and thus allows ozone concentration to increase. When adsorbed HOX are desorbed and dissociate into other by‐products, HOX are not immediately available to destroy ozone. This results in larger ozone concentrations than if desorbed HOX are released directly back into their gaseous states. When using the hybrid scheme, ozone TCA is increased up to 50\% where the ozone deficit is greatest, demonstrating the best agreement with observations, and implying that HOX radicals are both retained when adsorbed and dissociate.},
	language = {en},
	number = {7},
%	urldate = {2025-06-17},
	journal = {Journal of Geophysical Research: Planets},
	author = {Brown, M. A. J. and Patel, M. R. and Lewis, S. R. and Holmes, J. A. and Lefèvre, F. and Mason, J. P. and Crismani, M.},
%	month = jul,
	year = {2024},
	pages = {e2024JE008302},
	file = {Brown et al. - 2024 - The Role and Lifetime of Dissociative Heterogeneou.pdf:C\:\\Users\\bbenne\\Zotero\\storage\\8KDLKXEX\\Brown et al. - 2024 - The Role and Lifetime of Dissociative Heterogeneou.pdf:application/pdf},
}

@article{benne_3d_2025,
	title = {A {3D} model simulation of hydrogen chloride photochemistry on {Mars}: {Comparison} with satellite data},
	copyright = {© 2025, ESO},
	issn = {0004-6361, 1432-0746},
	shorttitle = {A {3D} model simulation of hydrogen chloride photochemistry on {Mars}},
%	url = {https://www.aanda.org/articles/aa/abs/forth/aa53872-25/aa53872-25.html},
	doi = {10.1051/0004-6361/202553872},
	abstract = {Hydrogen chloride (HCl) was independently detected in the Martian atmosphere by the Nadir and Occultation for MArs Discovery (NOMAD) and Atmospheric Chemistry Suite (ACS) spectrometers aboard the ExoMars Trace Gas Orbiter (TGO). Photochemical models show that using gas-phase chemistry alone is insufficient to reproduce these data. Recent work has developed a heterogeneous chemical network within a 1D photochemistry model, guided by the seasonal variability in HCl. This variability includes detection almost exclusively during the dust season, a positive correlation with water vapour, and an anticorrelation with water ice. The aim of this work is to show that incorporating heterogeneous chlorine chemistry into a global 3D model of Martian photochemistry with conventional gas-phase chemistry can reproduce spatial and temporal changes in hydrogen chloride on Mars, as observed by instruments aboard the TGO. We incorporated this heterogeneous chlorine scheme into the Mars Planetary Climate Model (MPCM). After some refinements to the scheme, mainly associated with it being employed in a 3D model, we used it to model chlorine photochemistry during Mars Years (MYs) 34 and 35. These two years provide contrasting dust scenarios, with MY 34 featuring a global dust storm. We also examined correlations in the model results between HCl and other key atmospheric quantities, as well as production and loss processes, to understand the impact of different factors driving changes in HCl. We find that the 3D model of Martian photochemistry using the proposed heterogeneous chemistry is consistent with the changes in HCl observed by ACS in MY 34 and MY 35, including detections and 70\% of non-detections. For the remaining 30\% of non-detections, model HCl is higher than the ACS detection limit due to biases associated with water vapour, dust, or water ice content at these locations. As with previous 1D model calculations, we find that heterogeneous chemistry is required to describe the loss of HCl, resulting in a lifetime of a few sols that is consistent with the observed seasonal variation in HCl. As a result of this proposed chemistry, modelled HCl is correlated with water vapour, airborne dust, and temperature, and anticorrelated with water ice. Our work shows that this chemical scheme enables the reproduction of aphelion detections in MY 35.},
	language = {en},
%	urldate = {2025-07-02},
	journal = {Astronomy \& Astrophysics},
	author = {Benne, Benjamin and Palmer, Paul I. and Taysum, Benjamin M. and Olsen, Kevin S. and Lefèvre, Franck},
%	month = jun,
	year = {2025},
	note = {Publisher: EDP Sciences},
}

@article{Andreotti2009,
  author    = {Andreotti, Bruno and Fourri{\`e}re, Antoine and Ould{-}Kaddour, Fouzia and Murray, Brad and Claudin, Philippe},
  title     = {Giant aeolian dune size determined by the average depth of the atmospheric boundary layer},
  journal   = {Nature},
  volume    = {457},
  number    = {7233},
  pages     = {1120--1123},
  year      = {2009},
  doi       = {10.1038/nature07787}
}

@article{Atri2023,
  author    = {Atri, Dimitra and Abdelmoneim, Nour and Dhuri, Dilip B. and Simoni, Matteo},
  title     = {Diurnal variation of the surface temperature of {Mars} with the {Emirates Mars Mission}: A comparison with {Curiosity} and {Perseverance} rover measurements},
  journal   = {Monthly Notices of the Royal Astronomical Society: Letters},
  volume    = {518},
  number    = {1},
  pages     = {L1--L6},
  year      = {2023},
  doi       = {10.1093/mnrasl/slac132}
}

@article{Banham2018,
  author    = {Banham, S. G. and Gupta, S. and Rubin, D. M. and Watkins, J. A. and Sumner, D. Y. and Edgett, K. S. and Grotzinger, J. P. and Lewis, K. W. and Edgar, L. A. and Stack{-}Morgan, K. M. and Barnes, R.},
  title     = {Ancient Martian aeolian processes and palaeomorphology reconstructed from the Stimson formation on the lower slope of Aeolis Mons, Gale crater, Mars},
  journal   = {Sedimentology},
  volume    = {65},
  number    = {4},
  pages     = {993--1042},
  year      = {2018},
  doi       = {10.1111/sed.12440}
}

@article{Banham2022,
  author    = {Banham, S. G. and Gupta, S. and Rubin, D. M. and Bedford, C. C. and Edgar, L. A. and Bryk, A. B. and Dietrich, W. E. and Fedo, C. M. and Williams, R. M. E. and Caravaca, G. and Barnes, R.},
  title     = {Evidence for fluctuating wind in shaping an ancient Martian dune field: The Stimson formation at the Greenheugh pediment, Gale crater},
  journal   = {Journal of Geophysical Research: Planets},
  volume    = {127},
  number    = {9},
  pages     = {e2021JE007023},
  year      = {2022},
  doi       = {10.1029/2021JE007023}
%  url       = {https://agupubs.onlinelibrary.wiley.com/doi/10.1029/2021JE007023}
}

@article{Banham2021,
  author    = {Banham, Steven G. and Gupta, Sanjeev and Rubin, David M. and Edgett, Kenneth S. and Barnes, Ron and Van Beek, Julie and Watkins, Jennifer A. and Edgar, Lauren A. and Fedo, Christopher M. and Williams, Rebecca M. and Stack, Kathryn M.},
  title     = {A rock record of complex aeolian bedforms in a {Hesperian} desert landscape: The {Stimson} formation as exposed in the {Murray} buttes, {Gale} crater, {Mars}},
  journal   = {Journal of Geophysical Research: Planets},
  volume    = {126},
  number    = {4},
  pages     = {e2020JE006554},
  year      = {2021},
  doi       = {10.1029/2020JE006554}
}

@article{Barrett2026,
  author    = {Barrett, A. M. and Balme, M. R. and Favaro, E. A. and Woods, M. J. and Malinowski, M. and Elsarky, L. and Rajendran, K. and Holmes, J. A. and Patel, M. R.},
  title     = {Detecting and measuring {Transverse Aeolian Ridges} on {Mars} using deep learning},
  journal   = {Icarus},
  pages     = {116947},
  year      = {2026},
  doi       = {10.1016/j.icarus.2024.116947}
}

@article{Battersby1997,
    title = {Mars rover meets rock with complex past},
    author = {Battersby, S.},
    journal = {Nature},
    volume = {388},
    number = {215},
    year = {1997},
    doi = {10.1038/40712}
}

@article{Berger2016,
    title = {A global {M}ars dust composition refined by the {A}lpha‐{P}article {X}‐ray {S}pectrometer in {G}ale {C}rater},
    author = {Berger, J.A. and Schmidt, M.E. and Gellert, R. and Campbell, J.L. and King, P.L. and Flemming, R.L. and Ming, D.W. and Clark, B.C. and Pradler, I. and VanBommel, S.J. and Minitti, M.E.},
    journal = {Geophysical Research Letters},
    volume = {43},
    number = {1},
    pages = {67--75},
    year = {2016}
}

@article{Campbell2025,
  author    = {Campbell, C. L. and Moores, J. E.},
  title     = {Using imagery from the Mars Science Laboratory mission to understand wind properties of aerosols aloft},
  journal   = {Planetary and Space Science},
  volume    = {263},
  pages     = {106119},
  year      = {2025},
  doi       = {10.1016/j.pss.2025.106119}
%  url       = {https://www.sciencedirect.com/science/article/pii/S0032063325000868}
}

@article{Chide2025,
  author    = {Chide, B. and Lorenz, R. D. and Montmessin, F. and Maurice, S. and Parot, Y. and Hueso, R. and Martinez, G. and Vicente{-}Retortillo, A. and Jacob, X. and Lemmon, M. and Dubois, B.},
  title     = {Detection of triboelectric discharges during dust events on {Mars}},
  journal   = {Nature},
  volume    = {647},
  number    = {8091},
  pages     = {865--869},
  year      = {2025},
%  url       = {https://www.nature.com/articles/s41586-025-09736-y},
  doi       = {10.1038/s41586-025-09736-y}
}

@article{Cooper1996,
  author    = {Cooper, P. L. and Abbatt, J. P. D.},
  title     = {Heterogeneous interactions of OH and HO$_2$ radicals with surfaces characteristic of atmospheric particulate matter},
  journal   = {The Journal of Physical Chemistry},
  year      = {1996},
  volume    = {100},
  number    = {6},
  pages     = {2249--2254},
  doi       = {10.1021/jp953050h}
}

@article{Crisp2003,
  author    = {Crisp, J. A. and Adler, M. and Matijevic, J. R. and Squyres, S. W. and Arvidson, R. E. and Kass, D. M.},
  title     = {Mars Exploration Rover mission},
  journal   = {Journal of Geophysical Research},
  volume    = {108},
  pages    = {E12},
  number     = {8061},
  year      = {2003},
  doi       = {10.1029/2002JE002038},
%  url       = {https://doi.org/10.1029/2002JE002038}
}

@article{DeSanctis2022,
  author    = {De Sanctis, Maria Cristina and Altieri, Francesca and Ammannito, Eleonora and De Angelis, Simone and Ehlmann, Bethany and Ferrari, Maurizio and Frigeri, Alessandro and Fonte, Silvia and Formisano, Mario and Giardino, Marco and Apuzzo, Alessandro},
  title     = {Exploring the shallow subsurface of {Mars} with the {Ma\_MISS} spectrometer on the {ExoMars} rover {Rosalind Franklin}},
  journal   = {The Planetary Science Journal},
  volume    = {3},
  number    = {6},
  pages     = {142},
  year      = {2022},
  doi       = {10.3847/PSJ/ac694f}
}

@article{Donahue1982,
  author    = {Donahue, T. M. and Hoffman, J. H. and Hodges, R. R. and Watson, A. J.},
  title     = {Venus was wet: a measurement of the ratio of deuterium to hydrogen},
  journal   = {Science},
  volume    = {216},
  pages     = {630--633},
  year      = {1982},
  doi       = {10.1126/science.216.4546.630}
}

@article{Fairen2010,
  author    = {Fair{\'e}n, A. G. and Davila, A. F. and Lim, D. and Bramall, N. and Bonaccorsi, R. and Zavaleta, J. and Uceda, E. R. and Stoker, C. and Wierzchos, J. and Dohm, J. M. and Amils, R.},
  title     = {Astrobiology through the ages of Mars: the study of terrestrial analogues to understand the habitability of Mars},
  journal   = {Astrobiology},
  volume    = {10},
  number    = {8},
  pages     = {821--843},
  year      = {2010},
  doi       = {10.1089/ast.2009.0440},
%  url       = {https://journals.sagepub.com/doi/full/10.1089/ast.2009.0440}
}

@inproceedings{Favaro2025,
  author    = {Favaro, E. A. and Patel, M. R. and Rajendran, K. and Holmes, J. A.},
  title     = {A {M}esoscale {M}odelling {A}pproach to {U}nderstanding {M}ars' {A}eolian {L}andscapes},
  booktitle = {EPSC-DPS Joint Meeting 2025},
  year      = {2025},
  number    = {EPSC-DPS2025-1583},
  publisher = {Copernicus Meetings},
}

@article{Fornaro2025,
    title = {Evidence for polycyclic aromatic hydrocarbons detected in sulfates at Jezero crater by the Perseverance rover},
    author = {Fornaro, T. and Sharma, S. and Jakubek, R.S. and Poggiali, G. and Brucato, J.R. and Bhartia, R. and Steele, A. and Murphy, A.E. and Tice, M.M. Schulte, M.D. and Hand, K.P.},
    journal = {Icarus},
    volume = {9},
    number = {11},
    pages = {1648-1661},
    year = {2025},
}

@article{Franz2015,
  author    = {Franz, H. B. and Trainer, M. G. and Wong, M. H. and Mahaffy, P. R. and Atreya, S. K. and Manning, H. L. and Stern, J. C.},
  title     = {Reevaluated Martian atmospheric mixing ratios from the mass spectrometer on the Curiosity rover},
  journal   = {Planetary and Space Science},
  volume    = {109},
  pages     = {154--158},
  year      = {2015},
  doi       = {10.1016/j.pss.2015.03.004},
%  url       = {https://www.sciencedirect.com/science/article/pii/S0032063315000495}
}

@article{Geissler2010,
  author    = {Geissler, P. E. and Sullivan, R. and Golombek, M. and Johnson, J. R. and Herkenhoff, K. and Bridges, N. and Vaughan, A. and Maki, J. and Parker, T. and Bell, J. F.},
  title     = {Gone with the wind: Eolian erasure of the {M}ars {R}over tracks},
  journal   = {Journal of Geophysical Research: Planets},
  volume    = {115},
  number    = {E7},
  year      = {2010},
  doi       = {10.1029/2010JE003674},
%  url       = {https://agupubs.onlinelibrary.wiley.com/doi/full/10.1029/2010JE003674}
}

@article{GomezElvira2014,
  author    = {G{\'o}mez-Elvira, J. and Armiens, C. and Carrasco, I. and Genzer, M. and G{\'o}mez, F. and Haberle, R. and Hamilton, V. E. and Harri, A. M. and Kahanp{\"a}{\"a}, H. and Kemppinen, O. and Lepinette, A.},
  title     = {Curiosity's Rover Environmental Monitoring Station: Overview of the First 100 Sols},
  journal   = {Journal of Geophysical Research: Planets},
  volume    = {119},
  number    = {7},
  pages     = {1680--1688},
  year      = {2014},
  doi       = {10.1002/2013JE004576},
%  url       = {https://doi.org/10.1002/2013JE004576}
}

@article{gonzalezgalindo2013,
    title = {Three‐dimensional {M}artian ionosphere model: {I}. {T}he photochemical ionosphere below 180 km},
    author = {González‐Galindo, F. and Chaufray, J.Y. and López‐Valverde, M.A. and Gilli, G. and Forget, F. and Leblanc, F. and Modolo, R. and Hess, S. and Yagi, M.},
    journal = {Journal of Geophysical Research: Planets},
    volume = {118},
    number = {10},
    pages = {2105--2123},
    year = {2013}
}

@article{Greeley2010,
  author    = {Greeley, R. and Waller, D. A. and Cabrol, N. A. and Landis, G. A. and Lemmon, M. T. and Neakrase, L. D. V. and Pendleton Hoffer, M. and Thompson, S. D. and Whelley, P. L.},
  title     = {Gusev Crater, Mars: Observations of three dust devil seasons},
  journal   = {Journal of Geophysical Research},
  volume    = {115},
  number    = {E00F02},
  year      = {2010},
  doi       = {10.1029/2010JE003608},
%  url       = {https://agupubs.onlinelibrary.wiley.com/doi/full/10.1029/2010JE003608}
}

@article{Greeley2003,
  author    = {Greeley, Ronald and Thompson, Sheryl D.},
  title     = {Mars: Aeolian features and wind predictions at the {Terra Meridiani} and {Isidis Planitia} potential {Mars Exploration Rover} landing sites},
  journal   = {Journal of Geophysical Research: Planets},
  volume    = {108},
  number    = {E12},
  year      = {2003},
  doi       = {10.1029/2003JE002110}
}

@inproceedings{Gregory2025,
  author    = {Gregory, B. and Olsen, K. and Millour, E. and Brown, M.},
  title     = {Modelling the {I}nfluence of {O}xidative {C}hemistry on {T}race {G}ases in {M}ars' {A}tmosphere},
  booktitle = {Proceedings of the EuroPlanet Science Congress},
  year      = {2025},
  number     = {EPSC-DPS2025-1687},
  organization = {Copernicus Meetings}
}

@article{Grenfell2022,
    author = {Grenfell, J.L. and Wunderlich, F. and Sinnhuber, M. and Herbst, K. and Lehmann, R. and Scheucher, M. and Gebauer, S. and Arnold, G. and Rauer, H.},
    title = {Atmospheric processes affecting methane on {M}ars},
    journal = {Icarus},
    year = {2022},
    volume = {382},
    pages = {114940},
    doi = {10.1016/j.icarus.2022.114940}
}

@article{Grotzinger2012,
  author    = {Grotzinger, J. P. and Crisp, J. and Vasavada, A. R. and Anderson, R. C. and Baker, C. J. and Barry, R. and Blake, D. F. and Conrad, P. and Edgett, K. S. and Ferdowski, B. and Gellert, R.},
  title     = {Mars Science Laboratory mission and science investigation},
  journal   = {Space Science Reviews},
  volume    = {170},
  number    = {1},
  pages     = {5--56},
  year      = {2012},
  doi       = {10.1007/s11214-012-9892-2},
%  url       = {https://link.springer.com/article/10.1007/s11214-012-9892-2}
}

@article{Hayes2011,
  author    = {Hayes, A. G. and Grotzinger, J. P. and Edgar, L. A. and Squyres, S. W. and Watters, W. A. and Sohl-Dickstein, J.},
  title     = {Reconstruction of eolian bed forms and paleocurrents from cross-bedded strata at Victoria Crater, Meridiani Planum, Mars},
  journal   = {Journal of Geophysical Research: Planets},
  volume    = {116},
  number    = {E7},
  year      = {2011},
  doi       = {10.1029/2010JE003688},
%  url       = {https://agupubs.onlinelibrary.wiley.com/doi/full/10.1029/2010JE003688}
}

@inproceedings{Helbert2025,
  author    = {Helbert, Jörn and Bründl, Tara-Maria and Haag, Martin and Lindner, Martin and Ordoubadian, Björn and Wittig, Sven and the L4 Expert Committee and the L4 Payload Working Group},
  title     = {The Mission to Enceladus -- The ESA L4 Mission},
  booktitle = {EPSC-DPS Joint Meeting 2025},
  series    = {EPSC Abstracts},
  volume    = {18},
  pages     = {EPSC-DPS2025-1307},
  year      = {2025},
  doi       = {10.5194/epsc-dps2025-1307},
%  url       = {https://doi.org/10.5194/epsc-dps2025-1307}
}

@article{Holmes2020,
    author = {Holmes, J.A. and Lewis and S.R. and Patel, M.R.},
    title = {Open{MARS}: {A} global record of martian weather from 1999 to 2015},
    year = {2020},
    journal = {Planetary and Space Science},
    volume = {188},
    pages = {104962}
}

@article{Jackson2020,
  author    = {Jackson, D. W. and Cooper, A. and Green, A. and Beyers, M. and Guisado-Pintado, E. and Wiles, E. and Benallack, K. and Balme, M.},
  title     = {Reversing transverse dunes: Modelling of airflow switching using 3D computational fluid dynamics},
  journal   = {Earth and Planetary Science Letters},
  volume    = {544},
  pages     = {116363},
  year      = {2020},
  doi       = {10.1016/j.epsl.2020.116363}
}

@article{Jakosky2018,
  author    = {Jakosky, B.M. and Brain, D. and Chaffin, M. and Curry, S. and Deighan, J. and Grebowsky, J. and Halekas, J. and Leblanc, F. and Lillis, R. and Luhmann, J.G. and Andersson, L. and N. Andre and D. Andrews and D. Baird and D. Baker and J. Bell and M. Benna and D. Bhattacharyya and S. Bougher and C. Bowers and P. Chamberlin and J.-Y. Chaufray and J. Clarke and G. Collinson and M. Combi and J. Connerney and K. Connour and J. Correira and K. Crabb and F. Crary and T. Cravens and M. Crismani and G. Delory and R. Dewey and G. DiBraccio and C. Dong and Y. Dong and P. Dunn and H. Egan and M. Elrod and S. England and F. Eparvier and R. Ergun and A. Eriksson and T. Esman and J. Espley and S. Evans and K. Fallows and X. Fang and M. Fillingim and C. Flynn and A. Fogle and C. Fowler and J. Fox and M. Fujimoto and P. Garnier and Z. Girazian and H. Groeller and J. Gruesbeck and O. Hamil and K.G. Hanley and T. Hara and Y. Harada and J. Hermann and M. Holmberg and G. Holsclaw and S. Houston and S. Inui and S. Jain and R. Jolitz and A. Kotova and T. Kuroda and D. Larson and Y. Lee and C. Lee and F. Lefevre and C. Lentz and D. Lo and R. Lugo and Y.-J. Ma and P. Mahaffy and M.L. Marquette and Y. Matsumoto and M. Mayyasi and C. Mazelle and W. McClintock and J. McFadden and A. Medvedev and M. Mendillo and K. Meziane and Z. Milby and D. Mitchell and R. Modolo and F. Montmessin and A. Nagy and H. Nakagawa and C. Narvaez and K. Olsen and D. Pawlowski and W. Peterson and A. Rahmati and K. Roeten and N. Romanelli and S. Ruhunusiri and C. Russell and S. Sakai and N. Schneider and K. Seki and R. Sharrar and S. Shaver and D.E. Siskind and M. Slipski and Y. Soobiah and M. Steckiewicz and M.H. Stevens and I. Stewart and A. Stiepen and S. Stone and V. Tenishev and N. Terada and K. Terada and E. Thiemann and R. Tolson and G. Toth and J. Trovato and M. Vogt and T. Weber and P. Withers and S. Xu and R. Yelle and E. Yiğit and R. Zurek},
  title     = {Loss of the {M}artian atmosphere to space: {P}resent-day loss rates determined from {MAVEN} observations and integrated loss through time},
  journal   = {Icarus},
  volume    = {315},
  pages     = {46-157},
  year      = {2018},
}

@article{Mellon2009,
  author    = {Mellon, Michael T. and Arvidson, Raymond E. and Sizemore, Hanna G. and Searls, Mary L. and Blaney, Diana L. and Cull, Sarah and Hecht, Michael H. and Heet, Todd L. and Keller, Horst U. and Lemmon, Mark T. and Markiewicz, W. J.},
  title     = {Ground ice at the {Phoenix} landing site: Stability state and origin},
  journal   = {Journal of Geophysical Research: Planets},
  volume    = {114},
  number    = {E1},
  year      = {2009},
  doi       = {10.1029/2009JE003417}
}

@incollection{Kahre2017,
    author = {Kahre, M.A. and Murphy, J.R. and Newman, C.E., and Wilson, R.J. and Cantor, B.A. and Lemmon, M.T. and Wolff, M.J.},
    title = {The {M}ars {D}ust {C}ycle},
    editor = {Haberle, R.M. and Clancy, R.T. and Forget, F. and Smith, M.D. and Zurek, R.W.},
    booktitle = {The {A}tmosphere and {C}limate of {M}ars},
    publisher = {Cambridge University Press},
    year = {2017},
    pages = {295--337},
    doi = {10.1017/9781139060172.010}
}

@article{Kammerer2022,
  author    = {Kammerer, J. and Quanz, S. P. and Dannert, F.},
  title     = {Large Interferometer For Exoplanets (LIFE)--VI. Detecting rocky exoplanets in the habitable zones of Sun-like stars},
  journal   = {Astronomy \& Astrophysics},
  volume    = {668},
  pages     = {A52},
  year      = {2022},
  doi       = {10.1051/0004-6361/202244249}
}

@article{Kite2014,
  author    = {Kite, E. S. and Williams, J. P. and Lucas, A. and Aharonson, O.},
  title     = {Low palaeopressure of the Martian atmosphere estimated from the size distribution of ancient craters},
  journal   = {Nature Geoscience},
  volume    = {7},
  number    = {5},
  pages     = {335--339},
  year      = {2014},
  doi       = {10.1038/ngeo2137},
%  url       = {https://www.nature.com/articles/ngeo2137}
}

@article{Kurokawa2018,
  author    = {Kurokawa, H. and Kurosawa, K. and Usui, T.},
  title     = {A lower limit of atmospheric pressure on early Mars inferred from nitrogen and argon isotopic compositions},
  journal   = {Icarus},
  volume    = {299},
  pages     = {443--459},
  year      = {2018},
  doi       = {10.1016/j.icarus.2016.12.003},
%  url       = {https://www.sciencedirect.com/science/article/pii/S0019103516303062}
}

@article{Lapotre2016,
  author    = {Lapotre, M. G. A. and Ewing, R. C. and Lamb, M. P. and Fischer, W. W. and Grotzinger, J. P. and Rubin, D. M. and Lewis, K. W. and Ballard, M. J. and Day, M. and Gupta, S. and Banham, S. G.},
  title     = {Large wind ripples on Mars: A record of atmospheric evolution},
  journal   = {Science},
  volume    = {353},
  number    = {6294},
  pages     = {55--58},
  year      = {2016},
  doi       = {10.1126/science.aaf3206},
%  url       = {https://www.science.org/doi/10.1126/science.aaf3206}
}

@incollection{Lefevre2017,
    author = {Lefèvre, F. and Krasnopolsky, V.},
    title = {Atmospheric photochemistry},
    editor = {Haberle, R.M. and Clancy, R.T. and Forget, F. and Smith, M.D. and Zurek, R.W.},
    booktitle = {The {A}tmosphere and {C}limate of {M}ars},
    publisher = {Cambridge University Press},
    year = {2017},
    pahes = {229--294}
}

@article{Lemmon2015,
  author    = {Lemmon, M. T. and Wolff, M. J. and Bell, J. F., III and Smith, M. D. and Cantor, B. A. and Smith, P. H.},
  title     = {Dust aerosol, clouds, and the atmospheric optical depth record over 5 Mars years of the Mars Exploration Rover mission},
  journal   = {Icarus},
  volume    = {251},
  pages     = {96--111},
  year      = {2015},
  doi       = {10.1016/j.icarus.2014.03.029},
%  url       = {https://www.sciencedirect.com/science/article/pii/S0019103514001559}
}

@article{Lemmon2024,
  author    = {Lemmon, M. T. and Vicente-Retortillo, A. and Guzewich, S. D. and Ju{\'a}rez, M. D. L. T. and Innanen, A. C. and Campbell, C. L. and Maki, J. N. and Malin, M. C. and Moores, J. E.},
  title     = {Iridescence reveals the formation and growth of ice aerosols in Martian noctilucent clouds},
  journal   = {Geophysical Research Letters},
  volume    = {51},
  number    = {23},
  pages     = {e2024GL111183},
  year      = {2024},
  doi       = {10.1029/2024GL111183},
%  url       = {https://agupubs.onlinelibrary.wiley.com/doi/full/10.1029/2024GL111183}
}

@article{Lewis1995,
    author = {Lewis, S.R. and Read, P.L.},
    title = {An operational data assimilation scheme for the {M}artian atmosphere},
    year = {1995},
    journal = {Advances in Space Research},
    volume = {16},
    number = {6},
    pages = {9--13}
}

@article{Lewis2007,
    author = {Lewis, S.R. and Read, P.L. and Conrath, B.J. and Pearl, J.C. and Smith, M.D.},
    title = {Assimilation of thermal emission spectrometer atmospheric data during the {M}ars {G}lobal {S}urveyor aerobraking period},
    year = {2007},
    journal = {Icarus},
    volume = {192},
    number = {2},
    pages = {327--347},
    doi = {10.1016/j.icarus.2007.08.009}
}

@article{Lorenz2014,
  author    = {Lorenz, R. D. and Bridges, N. T. and Rosenthal, A. A. and Donkor, E.},
  title     = {Elevation dependence of bedform wavelength on Tharsis Montes, Mars: Atmospheric density as a controlling parameter},
  journal   = {Icarus},
  volume    = {230},
  pages     = {77--80},
  year      = {2014},
  doi       = {10.1016/j.icarus.2013.07.011}
}

@article{Mangold2021,
  author    = {Mangold, N. and et al.},
  title     = {Perseverance rover reveals an ancient delta-lake system and flood deposits at Jezero crater, Mars},
  journal   = {Science},
  volume    = {374},
  pages     = {711--717},
  year      = {2021},
  doi       = {10.1126/science.abl4051},
%  url       = {https://doi.org/10.1126/science.abl4051}
}

@article{McCulloch2023,
    author = {McCulloch, D. and Sergeev, D.E. and Mayne, N. and Bate, M. and Manners, J. and Boutle, I. and Drummond, B. and Kohary, K.},
    title = {A modern-day {M}ars climate in the {M}et {O}ffice {U}nified {M}odel: dry simulations},
    year = {2023},
    journal = {Geoscientific Model Development},
    volume = {16},
    number = {2},
    pages = {621--657}
}

@article{McSween2000,
    author = {McSween Jr, H.Y. and Keil, K.},
    title = {Mixing relationships in the {M}artian regolith and the composition of globally homogeneous dust},
    year = {2000},
    journal = {Geochimica et Cosmochimica Acta},
    volume = {64},
    number = {12},
    pages = {2155-2166}
}

@article{Miles2023,
  author    = {Miles, B. E. and Biller, B. A. and Patapis, P. and Worthen, K. and Rickman, E. and Hoch, K. K. and Skemer, A. and Perrin, M. D. and Whiteford, N. and Chen, C. H. and Sargent, B.},
  title     = {The JWST Early-release Science Program for Direct Observations of Exoplanetary Systems II: A 1 to 20 $\mu$m Spectrum of the Planetary-mass Companion VHS~1256--1257~b},
  journal   = {The Astrophysical Journal Letters},
  volume    = {946},
  number    = {1},
  pages     = {L6},
  year      = {2023},
  doi       = {10.3847/2041-8213/acbbc8}
}

@article{Mondro2025,
  author    = {Mondro, C. A. and Fedo, C. M. and Grotzinger, J. P. and Lamb, M. P. and Gupta, S. and Dietrich, W. E. and Banham, S. and Weitz, C. M. and Gasda, P. and Edgar, L. A. and Rubin, D.},
  title     = {Wave ripples formed in ancient, ice-free lakes in Gale crater, Mars},
  journal   = {Science Advances},
  volume    = {11},
  number    = {3},
  pages     = {eadr0010},
  year      = {2025},
  doi       = {10.1126/sciadv.adr0010},
%  url       = {https://www.science.org/doi/10.1126/sciadv.adr0010}
}

@article{Mulholland2013,
    author = {Mulholland, D.P. and Read, P.L. and Lewis, S.R.},
    title = {Simulating the interannual variability of major dust storms on Mars using variable lifting thresholds},
    year = {2013},
    journal = {Icarus},
    volume = {223},
    number = {1},
    pages = {344--358},
    doi = {10.1016/j.icarus.2012.12.003}
}

@article{Murphy2016,
  author    = {Murphy, J. and Steakley, K. and Balme, M. and Deprez, G. and Esposito, F. and Kahanp{\"a}{\"a}, H. and Lemmon, M. and Lorenz, R. and Murdoch, N. and Neakrase, L. and Patel, M.},
  title     = {Field measurements of terrestrial and Martian dust devils},
  journal   = {Space Science Reviews},
  volume    = {203},
  number    = {1},
  pages     = {39--87},
  year      = {2016},
  doi       = {10.1007/s11214-016-0283-y},
%  url       = {https://link.springer.com/article/10.1007/s11214-016-0283-y}
}

@article{Newman2017,
  author    = {Newman, C. E. and G{\'o}mez-Elvira, J. and Marin, M. and Navarro, S. and Torres, J. and Richardson, M. I. and Battalio, J. M. and Guzewich, S. D. and Sullivan, R. and de la Torre, M. and Vasavada, A. R.},
  title     = {Winds measured by the Rover Environmental Monitoring Station (REMS) during the Mars Science Laboratory (MSL) rover's Bagnold Dunes Campaign and comparison with numerical modeling using MarsWRF},
  journal   = {Icarus},
  volume    = {291},
  pages     = {203--231},
  year      = {2017},
  doi       = {10.1016/j.icarus.2016.12.016},
%  url       = {https://doi.org/10.1016/j.icarus.2016.12.016}
}

@article{Newman2002,
    author = {Newman, C.E. and Lewis, S.R. and Read, P.L. and Forget, F.},
    title = {Modeling the {M}artian dust cycle, 1. {R}epresentations of dust transport processes},
    year = {2002},
    journal = {Journal of Geophysical Research: Planets},
    volume = {107},
    number = {E12},
    pages = {6-1--6-18},
    doi = {10.1029/2002JE001910}
}

@article{OConnellCooper2018,
    author = {O'Connell-Cooper, C.D. and Thompson, L.M. and Spray, J.G. and Berger, J.A. and VanBommel, S.J. and Gellert, R. and Boyd, N.I. and DeSouza, E.},
    title = {Chemical diversity of sands within the linear and barchan dunes of the {B}agnold {D}unes, {G}ale {C}rater, as revealed by {APXS} onboard {C}uriosity},
    year = {2018},
    journal = {Geophysical Research Letters},
    volume = {45},
    number = {18},
    pages = {9460-9470},
    doi = {10.1029/2018GL079026}
}

@article{Ojha2018,
    author = {Ojha, L. and Lewis, K. and Karunatillake, S. and Schmidt, M.},
    title = {The {M}edusae {F}ossae {F}ormation as the single largest source of dust on {M}ars},
    year = {2018},
    journal = {Nature Communications},
    volume = {9},
    number = {1},
    pages = {2867},
}

@article{Owen1988,
  author    = {Owen, T. and Maillard, J. P. and de Bergh, C. and Lutz, B. L.},
  title     = {Deuterium on {Mars}: The abundance of HDO and the value of D/H},
  journal   = {Science},
  volume    = {240},
  number    = {4860},
  pages     = {1767--1770},
  year      = {1988},
  doi       = {10.1126/science.240.4860.1767}
}

@article{Padovani2023,
  author    = {Padovani, P. and Cirasuolo, M.},
  title     = {The Extremely Large Telescope},
  journal   = {Contemporary Physics},
  volume    = {64},
  number    = {1},
  pages     = {47--64},
  year      = {2023},
  doi       = {10.1080/00107514.2023.2266921},
%  url       = {https://doi.org/10.1080/00107514.2023.2266921}
}

@article{PlaGarcia2019,
  author    = {Pla-Garcia, J. and Rafkin, S. C. and Karatekin, {\O}. and Gloesener, E.},
  title     = {Comparing MSL Curiosity rover TLS-SAM methane measurements with Mars Regional Atmospheric Modeling System atmospheric transport experiments},
  journal   = {Journal of Geophysical Research: Planets},
  volume    = {124},
  number    = {8},
  pages     = {2141--2167},
  year      = {2019},
  doi       = {10.1029/2018JE005824},
%  url       = {https://agupubs.onlinelibrary.wiley.com/doi/full/10.1029/2018JE005824}
}

@article{Ramisetty2018,
    author = {Ramisetty, R. and Abdelmonem, A. and Shen, X. and Saathoff, H. and Leisner, T. and Mohr, C.},
    title = {Exploring femtosecond laser ablation in single-particle aerosol mass spectrometry},
    year = {2018},
    journal = {Atmospheric Measurement Techniques},
    volume = {11},
    number = {7},
    pages = {4345-4360},
    doi = {10.5194/amt-11-4345-2018}
}

@article{Roberts2025,
  author    = {Roberts, A. L. and Gupta, S. and Banham, S. G. and Cowart, A. and Edgar, L. A. and Rapin, W. and Dietrich, W. E. and Davis, J. M. and Kite, E. S. and Caravaca, G. and Mondro, C. A.},
  title     = {Paleo{-}scours within the layered sulfate{-}bearing unit at Gale Crater, {Mars}: Evidence for intense wind erosion},
  journal   = {Journal of Geophysical Research: Planets},
  volume    = {130},
  number    = {5},
  pages     = {e2024JE008680},
  year      = {2025},
  doi       = {10.1029/2024JE008680}
}

@article{Rodriguez2018,
  author    = {Rodriguez, S. and Le Mou{\'e}lic, S. and Barnes, J. W. and Kok, J. F. and Rafkin, S. C. R. and Lorenz, R. D. and Charnay, B. and Radebaugh, J. and Narteau, C. and Cornet, T. and Bourgeois, O.},
  title     = {Observational evidence for active dust storms on {Titan} at equinox},
  journal   = {Nature Geoscience},
  volume    = {11},
  number    = {10},
  pages     = {727--732},
  year      = {2018},
  doi       = {10.1038/s41561-018-0233-2}
}

@article{Rubin1987,
  author    = {Rubin, D. M. and Hunter, R. E.},
  title     = {Bedform alignment in directionally varying flows},
  journal   = {Science},
  volume    = {237},
  number    = {4812},
  pages     = {276--278},
  year      = {1987},
  doi       = {10.1126/science.237.4812.276},
%  url       = {https://www.science.org/doi/10.1126/science.237.4812.276}
}

@article{Rubin2012,
  author    = {Rubin, D. M.},
  title     = {A unifying model for planform straightness of ripples and dunes in air and water},
  journal   = {Earth-Science Reviews},
  volume    = {113},
  number    = {3--4},
  pages     = {176--185},
  year      = {2012},
  doi       = {10.1016/j.earscirev.2012.03.002},
%  url       = {https://www.sciencedirect.com/science/article/pii/S0012825212000475}
}

@article{Smith2008,
    author = {Smith, M.D.},
    title = {Spacecraft observations of the {M}artian atmosphere},
    year = {2008},
    journal = {Annu. Rev. Earth Planet. Sci.},
    volume = {36},
    number = {1},
    pages = {191--219},
}

@article{Smith2006,
  author    = {Smith, M. D. and Wolff, M. J. and Spanovich, N. and Ghosh, A. and Banfield, D. and Christensen, P. R. and Landis, G. A. and Squyres, S. W.},
  title     = {One {M}artian year of atmospheric observations using {MER} {M}ini-{TES}},
  journal   = {Journal of Geophysical Research: Planets},
  volume    = {111},
  number    = {E12},
  year      = {2006},
  doi       = {10.1029/2006JE002770},
%  url       = {https://doi.org/10.1029/2006JE002770}
}

@article{Smith2016,
  author    = {Smith, M. D. and Zorzano, M. P. and Lemmon, M. and Mart{\'i}n-Torres, J. and de Cal, T. M.},
  title     = {Aerosol optical depth as observed by the Mars Science Laboratory REMS UV photodiodes},
  journal   = {Icarus},
  volume    = {280},
  pages     = {234--248},
  year      = {2016},
  doi       = {10.1016/j.icarus.2016.07.007},
%  url       = {https://www.sciencedirect.com/science/article/pii/S0019103516304158}
}

@article{Spiga2018,
  author    = {Spiga, A. and Banfield, D. and Teanby, N. A. and Forget, F. and Lucas, A. and Kenda, B. and Rodriguez Manfredi, J. A. and Widmer-Schnidrig, R. and Murdoch, N. and Lemmon, M. T. and Garcia, R. F.},
  title     = {Atmospheric science with InSight},
  journal   = {Space Science Reviews},
  volume    = {214},
  number    = {7},
  pages     = {109},
  year      = {2018},
  doi       = {10.1007/s11214-018-0543-0},
%  url       = {https://link.springer.com/article/10.1007/s11214-018-0543-0}
}

@article{Spohn2022,
  author    = {Spohn, Tilman and Hudson, Thomas L. and Witte, Lars and Wippermann, Tobias and Wisniewski, Lars and Kedziora, Bjoern and Vrettos, Christos and Lorenz, Ralph D. and Golombek, Matthew and Lichtenheldt, Roy and Grott, Matthias},
  title     = {The {InSight{-}HP3} mole on {Mars}: Lessons learned from attempts to penetrate to depth in the {Martian} soil},
  journal   = {Advances in Space Research},
  volume    = {69},
  number    = {8},
  pages     = {3140--3163},
  year      = {2022},
  doi       = {10.1016/j.asr.2022.02.024}
}

@inproceedings{Stella2010,
  author    = {Stella, P. M. and Herman, J. A.},
  title     = {The Mars surface environment and solar array performance},
  booktitle = {2010 35th {IEEE} {P}hotovoltaic {S}pecialists {C}onference},
  pages     = {2631--2635},
  year      = {2010},
  publisher = {IEEE},
  doi       = {10.1109/PVSC.2010.5617185},
%  url       = {https://ieeexplore.ieee.org/document/5617185}
}

@article{Streeter2020,
    author = {Streeter, P.M. and Lewis, S.R. and Patel, M.R. and Holmes, J.A. and Kass, D.M.},
    title = {Surface warming during the 2018/{M}ars year 34 global dust storm},
    year = {2020},
    journal = {Geophysical Research Letters},
    volume = {47},
    number = {9},
    pages = {e2019GL083936},
}

@article{Sullivan2000,
  author    = {Sullivan, R. and Greeley, R. and Kraft, M. and Wilson, G. and Golombek, M. and Herkenhoff, K. and Murphy, J. and Smith, P.},
  title     = {Results of the Imager for Mars Pathfinder windsock experiment},
  journal   = {Journal of Geophysical Research: Planets},
  volume    = {105},
  number    = {E10},
  pages     = {24547--24562},
  year      = {2000},
  doi       = {10.1029/1999JE001234},
%  url       = {https://doi.org/10.1029/1999JE001234}
}

@article{Temel2021,
  author    = {Temel, Onur and Karatekin, {\"O}zg{\"u}r and Mischna, Michael A. and Senel, Cihan B. and Mart{\'\i}nez, Germ{\'a}n and Gloesener, Elia and Van Hoolst, Tim},
  title     = {Strong seasonal and regional variations in the evaporation rate of liquid water on {Mars}},
  journal   = {Journal of Geophysical Research: Planets},
  volume    = {126},
  number    = {10},
  pages     = {e2021JE006867},
  year      = {2021},
  doi       = {10.1029/2021JE006867}
}

@article{Viudez-Moreiras2025,
    author = {Viúdez-Moreiras, D. and Smith, M.D. and Wolff, M. and Brown, M.A. and Daerden, F. and Zorzano, M.P. and Apestigue, V. and Arruego, I. and García, E. and Jiménez, J.J. and Toledo, D.},
    title = {Evidence of an enhanced near-surface ozone layer at tropical latitudes on {M}ars},
    year = {2025},
    journal = {Proceedings of the National Academy of Sciences},
    volume = {122},
    number = {48},
    pages = {e2511744122}
}

@article{Weitz2022,
    author = {Weitz, C.M. and O’Connell‐Cooper, C. and Thompson, L. and Sullivan, R.J. and Baker, M. and Grant, J.A.},
    title = {The physical properties and geochemistry of grains on aeolian bedforms at {G]ale {C}rater, {M}ars}},
    year = {2022},
    journal = {Journal of Geophysical Research: Planets},
    volume = {127},
    number = {11},
    pages = {e2021JE007061},
    doi = {10.1029/2021JE007061}
}

@article{Williams2013,
    author = {Williams, R.M. and Irwin III, R.P. and Burr, D.M. and Harrison, T. and McClelland, P.},
    title = {Variability in martian sinuous ridge form: {C}ase study of {A}eolis {S}erpens in the {A}eolis {D}orsa, {M}ars, and insight from the {M}irackina paleoriver, {S}outh {A}ustralia},
    year = {2013},
    journal = {Icarus},
    volume = {225},
    number = {1},
    pages = {308-324},
}

@article{Wordsworth2016,
  author    = {Wordsworth, R. D.},
  title     = {The climate of early Mars},
  journal   = {Annual Review of Earth and Planetary Sciences},
  volume    = {44},
  pages     = {381--408},
  year      = {2016},
  doi       = {10.1146/annurev-earth-060115-012355},
%  url       = {https://www.annualreviews.org/content/journals/10.1146/annurev-earth-060115-012355}
}

@article{Wordsworth2017,
  author    = {Wordsworth, R. and Kalugina, Y. and Lokshtanov, S. and Vigasin, A. and Ehlmann, B. and Head, J. and Sanders, C. and Wang, H.},
  title     = {Transient reducing greenhouse warming on early {Mars}},
  journal   = {Geophysical Research Letters},
  volume    = {44},
  number    = {2},
  pages     = {665--671},
  year      = {2017},
  doi       = {10.1002/2016GL071766}
}

@article{Wu2026,
    author = {Wu, S. and Xie, L.},
    title = {Fluid threshold of windblown sand and dust on {M}ars considering electric field and insolation},
    year = {2026},
    journal = {Journal of Geophysical Research: Planets},
    volume = {131},
    pages = {e2025JE009046},
    doi = {10.1029/2025JE009046}
}

@article{Newman2022,
doi = {10.1126/sciadv.abn3783},
author = {Claire E. Newman  and Ricardo Hueso  and Mark T. Lemmon  and Asier Munguira  and Álvaro Vicente-Retortillo  and Víctor Apestigue  and Germán M. Martínez  and Daniel Toledo  and Rob Sullivan  and Ken E. Herkenhoff  and Manuel de la Torre Juárez  and Mark I. Richardson  and Alexander E. Stott  and Naomi Murdoch  and Agustín Sanchez-Lavega  and Michael J. Wolff  and Ignacio Arruego  and Eduardo Sebastián  and Sara Navarro  and Javier Gómez-Elvira  and Leslie Tamppari  and Michael D. Smith  and Alain Lepinette  and Daniel Viúdez-Moreiras  and Ari-Matti Harri  and Maria Genzer  and Maria Hieta  and Ralph D. Lorenz  and Pan Conrad  and Felipe Gómez  and Timothy H. McConnochie  and David Mimoun  and Christian Tate  and Tanguy Bertrand  and James F. Bell  and Justin N. Maki  and Jose Antonio Rodriguez-Manfredi  and Roger C. Wiens  and Baptiste Chide  and Sylvestre Maurice  and Maria-Paz Zorzano  and Luis Mora  and Mariah M. Baker  and Don Banfield  and Jorge Pla-Garcia  and Olivier Beyssac  and Adrian Brown  and Ben Clark  and Franck Montmessin  and Erik Fischer  and Priyaben Patel  and Teresa del Río-Gaztelurrutia  and Thierry Fouchet  and Raymond Francis  and Scott D. Guzewich },
title = {The dynamic atmospheric and aeolian environment of Jezero crater, Mars},
journal = {Science Advances},
volume = {8},
number = {21},
pages = {eabn3783},
year = {2022},
doi = {10.1126/sciadv.abn3783},
URL = {https://www.science.org/doi/abs/10.1126/sciadv.abn3783},
eprint = {https://www.science.org/doi/pdf/10.1126/sciadv.abn3783}}

@article{Baker2020,
author = {Baker, M. and Newman, C. and Charalambous, C. and Golombek, M. and Spiga, A. and Banfield, D. and Lemmon, M. and Banks, M. and Lorenz, R. and Garvin, J. and Grant, J. and Lewis, K. and Ansan, V. and Warner, N. and Weitz, C. and Wilson, S. and Rodriguez, S.},
title = {Vortex-Dominated Aeolian Activity at InSight's Landing Site, Part 2: Local Meteorology, Transport Dynamics, and Model Analysis},
journal = {Journal of Geophysical Research: Planets},
volume = {126},
number = {4},
pages = {e2020JE006514},
doi = {https://doi-org.eux.idm.oclc.org/10.1029/2020JE006514},
url = {https://agupubs-onlinelibrary-wiley-com.eux.idm.oclc.org/doi/abs/10.1029/2020JE006514},
eprint = {https://agupubs-onlinelibrary-wiley-com.eux.idm.oclc.org/doi/pdf/10.1029/2020JE006514},
note = {e2020JE006514 2020JE006514},
year = {2021}
}

@article{VincenteRetortillo2023,
author = {Vicente-Retortillo, A. and Martínez, G. M. and Lemmon, M. T. and Hueso, R. and Johnson, J. R. and Sullivan, R. and Newman, C. E. and Sebastián, E. and Toledo, D. and Apéstigue, V. and Arruego, I. and Munguira, A. and Sánchez-Lavega, A. and Murdoch, N. and Gillier, M. and Stott, A. and Mora-Sotomayor, L. and Bertrand, T. and Tamppari, L. K. and Juárez, M. de la Torre and Rodríguez-Manfredi, J.-A.},
title = {Dust Lifting Through Surface Albedo Changes at Jezero Crater, Mars},
journal = {Journal of Geophysical Research: Planets},
volume = {128},
number = {4},
pages = {e2022JE007672},
doi = {https://doi-org.eux.idm.oclc.org/10.1029/2022JE007672},
url = {https://agupubs-onlinelibrary-wiley-com.eux.idm.oclc.org/doi/abs/10.1029/2022JE007672},
eprint = {https://agupubs-onlinelibrary-wiley-com.eux.idm.oclc.org/doi/pdf/10.1029/2022JE007672},
note = {e2022JE007672 2022JE007672},
year = {2023}
}

@article{Guzewich2023,
author = {Guzewich, Scott D. and Mason, Emily L. and Lemmon, Mark T. and Newman, Claire E. and Lewis, Kevin W.},
title = {Dust Lifting Observations With the Mars Science Laboratory Navigation Cameras},
journal = {Journal of Geophysical Research: Planets},
volume = {128},
number = {10},
pages = {e2023JE007959},
doi = {https://doi-org.eux.idm.oclc.org/10.1029/2023JE007959},
url = {https://agupubs-onlinelibrary-wiley-com.eux.idm.oclc.org/doi/abs/10.1029/2023JE007959},
eprint = {https://agupubs-onlinelibrary-wiley-com.eux.idm.oclc.org/doi/pdf/10.1029/2023JE007959},
note = {e2023JE007959 2023JE007959},
year = {2023}
}

@article{RodriguezManfredi2021,
	title = {The Mars Environmental Dynamics Analyzer, {MEDA}. A Suite of Environmental Sensors for the Mars 2020 Mission},
	volume = {217},
	issn = {1572-9672},
	url = {https://doi.org/10.1007/s11214-021-00816-9},
	doi = {10.1007/s11214-021-00816-9},
	pages = {48},
	number = {3},
	journaltitle = {Space Science Reviews},
	shortjournal = {Space Sci Rev},
	author = {Rodriguez-Manfredi, J. A. and de la Torre Juárez, M. and Alonso, A. and Apéstigue, V. and Arruego, I. and Atienza, T. and Banfield, D. and Boland, J. and Carrera, M. A. and Castañer, L. and Ceballos, J. and Chen-Chen, H. and Cobos, A. and Conrad, P. G. and Cordoba, E. and del Río-Gaztelurrutia, T. and de Vicente-Retortillo, A. and Domínguez-Pumar, M. and Espejo, S. and Fairen, A. G. and Fernández-Palma, A. and Ferrándiz, R. and Ferri, F. and Fischer, E. and García-Manchado, A. and García-Villadangos, M. and Genzer, M. and Giménez, S. and Gómez-Elvira, J. and Gómez, F. and Guzewich, S. D. and Harri, A.-M. and Hernández, C. D. and Hieta, M. and Hueso, R. and Jaakonaho, I. and Jiménez, J. J. and Jiménez, V. and Larman, A. and Leiter, R. and Lepinette, A. and Lemmon, M. T. and López, G. and Madsen, S. N. and Mäkinen, T. and Marín, M. and Martín-Soler, J. and Martínez, G. and Molina, A. and Mora-Sotomayor, L. and Moreno-Álvarez, J. F. and Navarro, S. and Newman, C. E. and Ortega, C. and Parrondo, M. C. and Peinado, V. and Peña, A. and Pérez-Grande, I. and Pérez-Hoyos, S. and Pla-García, J. and Polkko, J. and Postigo, M. and Prieto-Ballesteros, O. and Rafkin, S. C. R. and Ramos, M. and Richardson, M. I. and Romeral, J. and Romero, C. and Runyon, K. D. and Saiz-Lopez, A. and Sánchez-Lavega, A. and Sard, I. and Schofield, J. T. and Sebastian, E. and Smith, M. D. and Sullivan, R. J. and Tamppari, L. K. and Thompson, A. D. and Toledo, D. and Torrero, F. and Torres, J. and Urquí, R. and Velasco, T. and Viúdez-Moreiras, D. and Zurita, S. and {The MEDA team}},
	urldate = {2026-03-17},
	date = {2021-04-13},
	langid = {english},
}

@article{Hueso2023,
author = {Hueso, R. and Newman, C. E. and del Río-Gaztelurrutia, T. and Munguira, A. and Sánchez-Lavega, A. and Toledo, D. and Apéstigue, V. and Arruego, I. and Vicente-Retortillo, A. and Martínez, G. and Lemmon, M. and Lorenz, R. and Richardson, M. and Viudez-Moreiras, D. and de la Torre-Juarez, M. and Rodríguez-Manfredi, J. A. and Tamppari, L. K. and Murdoch, N. and Navarro-López, S. and Gómez-Elvira, J. and Baker, M. and Pla-García, J. and Harri, A. M. and Hieta, M. and Genzer, M. and Polkko, J. and Jaakonaho, I. and Makinen, T. and Stott, A. and Mimoun, D. and Chide, B. and Sebastian, E. and Banfield, D. and Lepinette-Malvite, A.},
title = {Convective Vortices and Dust Devils Detected and Characterized by Mars 2020},
journal = {Journal of Geophysical Research: Planets},
volume = {128},
number = {2},
pages = {e2022JE007516},
doi = {https://doi-org.eux.idm.oclc.org/10.1029/2022JE007516},
url = {https://agupubs-onlinelibrary-wiley-com.eux.idm.oclc.org/doi/abs/10.1029/2022JE007516},
eprint = {https://agupubs-onlinelibrary-wiley-com.eux.idm.oclc.org/doi/pdf/10.1029/2022JE007516},
note = {e2022JE007516 2022JE007516},
year = {2023}
}

@article{Lemmon2022,
author = {Lemmon, M. T. and Smith, M. D. and Viudez-Moreiras, D. and de la Torre-Juarez, M. and Vicente-Retortillo, A. and Munguira, A. and Sanchez-Lavega, A. and Hueso, R. and Martinez, G. and Chide, B. and Sullivan, R. and Toledo, D. and Tamppari, L. and Bertrand, T. and Bell III, J. F. and Newman, C. and Baker, M. and Banfield, D. and Rodriguez-Manfredi, J. A. and Maki, J. N. and Apestigue, V.},
title = {Dust, Sand, and Winds Within an Active Martian Storm in Jezero Crater},
journal = {Geophysical Research Letters},
volume = {49},
number = {17},
pages = {e2022GL100126},
doi = {https://doi-org.eux.idm.oclc.org/10.1029/2022GL100126},
url = {https://agupubs-onlinelibrary-wiley-com.eux.idm.oclc.org/doi/abs/10.1029/2022GL100126},
eprint = {https://agupubs-onlinelibrary-wiley-com.eux.idm.oclc.org/doi/pdf/10.1029/2022GL100126},
note = {e2022GL100126 2022GL100126},
year = {2022}
}

\end{document}